\documentclass[longauth]{aa}
\usepackage{subcaption}
\usepackage{graphicx}
\usepackage{xcolor}

\newcommand\hdist{2}

\usepackage{txfonts}
\usepackage{hyperref}
\usepackage{xspace}
\hypersetup{
    colorlinks=true,
    linkcolor=blue,
    citecolor=blue,
    filecolor=magenta,      
    urlcolor=cyan,
    pdftitle={Overleaf Example},
    pdfpagemode=FullScreen,
    }
\makeatletter
\renewcommand*\aa@pageof{, page \thepage{} of \pageref*{LastPage}}
\makeatother

\newcommand{\sfr}[2]{\ensuremath{\mathrm{SFR_{#1,#2}}}\xspace}
\newcommand{\cont}{{\ensuremath{\mathrm{cont}}}\xspace}
\newcommand{\sfrcten}{\sfr{\cont}{10}}
\newcommand{\sfrcninety}{\sfr{\cont}{90}}
\newcommand{\sfrchund}{\sfr{\cont}{100}}
\newcommand{\sfrnebten}{\sfr{neb}{10}}

\newcommand{\txn}[1]{\textnormal{#1}}
\newcommand{\gtsim}{\mbox{{\raisebox{-0.4ex}{$\stackrel{>}{{\scriptstyle\sim}}
$}}}}

\newcommand{\ssim}{\ensuremath{\sim\!}\xspace}

\newcommand{\M}{\hbox{$\txn{M}$}}

\newcommand{\Mstar}{\ensuremath{M_\star}\xspace}

\newcommand{\MSun}{\ensuremath{{\rm M}_\odot}\xspace}

\let\oldAA\AA
\renewcommand{\AA}{\oldAA\xspace}

\newcommand{\targetid}{JADES-GS-z7-01-QU\xspace}
\newcommand{\fesc}{\ensuremath{f_{\rm esc}}\xspace}

\newcommand{\LymanAlpha}{\text{Ly$\alpha$}\xspace}
\newcommand{\Halpha}{H$\alpha$\xspace}
\newcommand{\Hbeta}{H$\beta$\xspace}
\newcommand{\Hgamma}{H$\gamma$\xspace}
\newcommand{\DeltaMS}{$\Delta_{MS}$\xspace}

\newcommand{\jwst}{\textit{JWST}\xspace}

\newcommand{\ppxf}{{\sc ppxf}\xspace}
\newcommand{\bagpipes}{{\sc bagpipes}\xspace}
\newcommand{\beagle}{{\sc beagle}\xspace}
\newcommand{\prospector}{{\sc prospector}\xspace}

\begin{document}

   \title{JADES: Differing assembly histories of galaxies - Observational evidence for bursty SFHs and (mini-)quenching in the first billion years of the Universe}
%\title{JADES: Observational evidence for bursty SFHs in high-z galaxies imprinted into the stellar populations}

    \titlerunning{JADES: stellar populations and bursty SFHs}
   % \subtitle{I. Overviewing the $\kappa$-mechanism}

\author{
Tobias J.\ Looser
\inst{1,2}\fnmsep\thanks{tjl54@cam.ac.uk}
\and
Francesco D’Eugenio
\inst{1}\fnmsep\inst{2}
\and
Roberto Maiolino
\inst{1}\fnmsep\inst{2}\fnmsep\inst{3}
\and
Sandro Tacchella
\inst{1}\fnmsep\inst{2}
\and
Mirko Curti
\inst{4}\fnmsep\inst{1}\fnmsep\inst{2}
\and
Santiago Arribas
\inst{5}
\and
William M. Baker
\inst{1}\fnmsep\inst{2}\fnmsep
\and
Stefi Baum
\inst{6}
\and
Nina Bonaventura
\inst{7}\fnmsep\inst{8}\fnmsep\inst{9}
\and
Kristan Boyett
\inst{10}\fnmsep\inst{11}
\and
Andrew J. Bunker
\inst{12}
\and
Stefano Carniani 
\inst{13}
\and
Stephane Charlot
\inst{14}
\and
Jacopo Chevallard
\inst{12}
\and
Emma Curtis-Lake
\inst{15}
\and
A. Lola Danhaive
\inst{1}\fnmsep\inst{2}
\and
Daniel J.\ Eisenstein 
\inst{16}
\and
Anna de Graaff
\inst{17}
\and
Kevin Hainline
\inst{9}
\and
Zhiyuan Ji 
\inst{9}
\and
Benjamin D.\ Johnson
\inst{16}
\and
Nimisha Kumari
\inst{18}
\and
Erica Nelson
\inst{19}
\and
Eleonora Parlanti
\inst{13}
\and
Hans-Walter Rix 
\inst{17}
\and
Brant Robertson 
\inst{20}
\and
Bruno Rodr\'iguez Del Pino 
\inst{5}
\and
Lester Sandles
\inst{1}\fnmsep\inst{2}
\and
Jan Scholtz
\inst{1}\fnmsep\inst{2}
\and
Renske Smit
\inst{21}
\and
Daniel P. Stark
\inst{18}
\and
Hannah \"Ubler 
\inst{1}\fnmsep\inst{2}
\and
Christina C. Williams 
\inst{22}
\and
Chris Willott
\inst{23}
\and
Joris Witstok
\inst{1}\fnmsep\inst{2}
}

\institute{
Kavli Institute for Cosmology, University of Cambridge, Madingley Road, Cambridge, 
CB3 0HA, UK\\
\and
Cavendish Laboratory, University of Cambridge, 19 JJ Thomson Avenue, Cambridge CB3 0HE, UK\\
\and
Department of Physics and Astronomy, University College London, Gower Street, London WC1E 6BT, UK\\
\and
European Southern Observatory, Karl-Schwarzschild-Strasse 2, D-85748 Garching bei Muenchen, Germany\\
\and
Centro de Astrobiolog\'ia (CAB), CSIC–INTA, Cra. de Ajalvir Km.~4, 28850- Torrej\'on de Ardoz, Madrid, Spain\\
\and
Department of Physics and Astronomy, University of Manitoba, Winnipeg, MB R3T 2N2, Canada\\
\and
Cosmic Dawn Center (DAWN), Copenhagen, Denmark\\
\and
Niels Bohr Institute, University of Copenhagen, Jagtvej 128, DK-2200, Copenhagen, Denmark\\
\and
Steward Observatory University of Arizona 933 N. Cherry Avenue Tucson AZ 85721 USA\\
\and
School of Physics, University of Melbourne, Parkville 3010, VIC, Australia\\
\and
ARC Centre of Excellence for All Sky Astrophysics in 3 Dimensions (ASTRO 3D), Australia\\
\and
Department of Physics, University of Oxford, Denys Wilkinson Building, Keble Road, Oxford OX1 3RH, UK\\
\and
Scuola Normale Superiore, Piazza dei Cavalieri 7, I-56126 Pisa, Italy\\
\and
Sorbonne Universit\'e, CNRS, UMR 7095, Institut d'Astrophysique de Paris, 98 bis bd Arago, 75014 Paris, France\\
\and
Centre for Astrophysics Research, Department of Physics, Astronomy and Mathematics, University of Hertfordshire, Hatfield AL10 9AB, UK\\
\and
Center for Astrophysics $|$ Harvard \& Smithsonian, 60 Garden St., Cambridge MA 02138 USA\\
\and
Max-Planck-Institut f\"ur Astronomie, K\"onigstuhl 17, D-69117, Heidelberg, Germany\\
\and
AURA for European Space Agency, Space Telescope Science Institute, 3700 San Martin Drive. Baltimore, MD, 21210\\
\and
Department for Astrophysical and Planetary Science, University of Colorado, Boulder, CO 80309, USA\\
\and
Department of Astronomy and Astrophysics University of California, Santa Cruz, 1156 High Street, Santa Cruz CA 96054, USA\\
\and
Astrophysics Research Institute, Liverpool John Moores University, 146 Brownlow Hill, Liverpool L3 5RF, UK\\
\and
NSF’s National Optical-Infrared Astronomy Research Laboratory, 950 North Cherry Avenue, Tucson, AZ 85719, USA\\
\and
NRC Herzberg, 5071 West Saanich Rd, Victoria, BC V9E 2E7, Canada\\
}

   \authorrunning{T.\ J.\ Looser et al.}
   \date{}

% \abstract{}{}{}{}{} 
% 5 {} token are mandatory
 
  \abstract
  % context heading (optional)
  % {} leave it empty if necessary  
%   {}
  % conclusions heading (optional), leave it empty if necessary 
%   {}
{
% NEW ABSTRACT
We use deep NIRSpec spectroscopic data from the JADES survey to derive the star formation histories (SFHs) of a sample of 200 galaxies at 0.6$<$z$<$11 and spanning stellar masses from $\rm 10^6$ to $\rm 10^{9.5}~M_\odot$.
We find that galaxies at high-redshift, galaxies
above the Main Sequence (MS) and low-mass galaxies tend to host younger stellar populations than their low-redshift, massive, and below the MS counterparts. Interestingly,
the correlation between age, M$_*$ and SFR existed even earlier than Cosmic Noon, out to the earliest cosmic epochs.
However, these trends have a large scatter.
Indeed, there are
 examples of young stellar populations also below the MS, indicating recent (bursty) star formation in evolved systems.
We explore further the burstiness of the SFHs by using the ratio between SFR averaged over the last 10~Myr and averaged between 10~Myr and 100~Myr before the epoch of observation (\sfrcten/\sfrcninety).
We find that
high-redshift and low-mass galaxies have particularly bursty SFHs, while more massive and lower-redshift systems evolve more steadily.
We also present the discovery of
another (mini-)quenched galaxy at z = 4.4 (in addition to the one at z=7.3 reported by Looser et al. 2023), which might be only temporarily quiescent as a consequence of the extremely bursty evolution. Finally, we also find a steady decline of dust reddening of the stellar population approaching the earliest cosmic epochs, although some dust reddening is still observed in some of the highest redshift and most star forming systems.
}


   \maketitle
%
%-------------------------------------------------------------------

\section{Introduction} \label{sec:intro}

Understanding the nature and characteristics of stellar populations and the assembly histories of their host galaxies is a key objective in modern astrophysics. Stellar populations offer a unique window into the early universe, providing insights into the formation and evolution of galaxies, as well as shedding light on the processes that shaped the cosmos as we know it today. 

The launch of the James Webb Space Telescope \citep[\jwst,][]{Gardner2006,Gardner+2023}, with its unparalleled capabilities, ushers in a new era of astronomical exploration, providing the opportunity to uncover the intricate characteristics of stellar populations in objects at redshifts previously unattainable. The spectroscopic capabilities of the Near Infrared Spectrograph \citep[NIRSpec,][]{jakobsen+2022} onboard \jwst and its high sensitivity, enable us to push to fainter, lower mass and more distant sources and measure their properties accurately. With \jwst, we can explore the early stages of galaxy formation and trace the evolution of stellar populations across cosmic time.

%\todo{More literature on stellar populations/ previous work: Labbé et al. (2005), Eyles et al. (2007), Lee et al. (2015), Bouwens et al. (2021)}

Stellar populations are a crucial tool to understand the assembly of galaxies, as their past star-formation histories (SFHs) are imprinted in their stellar record. Hence, they encode valuable information about the various physical mechanisms which shaped their past star formation activity, such as stellar winds, feedback from supernovae (SN) and Active Galactic Nuclei (AGN), interactions and mergers, or environment. It is now widely thought, as predicted by numerical simulations \citep{Kawata&Gibson2003,Ceverino2021,Dome2023} and emerging observational evidence \cite{Caputi2007,Smit+2014,Smit+2015,Diaz_Santos2017,Endsley+2021,Endsley2022}, that these mechanisms can make SFHs "bursty", particularly in the early Universe \citep{Faucher-Giguere2018,Tacchella2016} and in low-mass systems \citep{Weisz2012, Tacchella2020}. Here "burstiness" refers to patterns of star formation characterized by episodic bursts of intense activity interspersed with "lull" phases, i.e. phases during which the galaxy is forming significantly fewer stars at the epoch of observation ($\sim$ 10 Myr time-scale) than during its recent past ($\sim$ 100 Myr time-scale), and potentially even quiescent periods, so-called mini-quenching events \citep{Dome2023,Gelli+2023}. Mini-quenching events refer to the state of a galaxy in which star formation is temporarily halted or strongly suppressed, i.e. $\mathrm{sSFR<0.2/t_{obs}}$, likely because the inflow of gas into the galaxy is disrupted, leading to a temporary halt of star formation activity. Unlike long-term quenching processes that lead to a permanent decline in star formation, mini-quenching events are transient and typically last for only a few tens to a hundred million years. 

Studying bursty SFHs and mini-quenching is therefore crucial for understanding the diversity of galaxy properties and the underlying physical processes that shape them. The timing, duration, and intensity of star formation bursts can influence the overall stellar mass assembly, the enrichment of heavy elements, and the morphological evolution of galaxies. Further, bursty SFHs are likely connected to the growth of supermassive black holes and the feedback mechanisms associated with AGN, even in low-mass systems \cite{Koudmani2019MNRAS.484.2047K}. Hence, the investigation of burstiness in SFHs of galaxies provides important constraints for theoretical models and is essential for constructing a comprehensive picture of galaxy evolution and unravelling the intricate processes that drive the diverse range of galaxy properties observed in the universe.

However, despite being thought to be a common phenomenon in galaxy evolution, observational evidence for bursty SFHs and mini-quenching is to date still sparse, mainly due to (pre-\jwst) instrumental limitations. Identifying and characterizing bursty SFHs requires high signal-to-noise (S/N) data, to disentangle the different star formation episodes within a galaxy. Observationally,
we (obviously) cannot know the future of any particular galaxy. Hence, it is difficult to assess on what time-scales, if not permanently, a particular galaxy remains quenched - making it observationally difficult to differentiate between mini-quenching and permanent-quenching on a galaxy-by-galaxy basis. Nonetheless, data on molecular cold gas in or around the galaxy, or evidence for infalling giant clouds of cold gas, could give important clues on whether the galaxy will obtain new fuel for star formation in the near future. Hence, unless such information is available for any particular galaxy we observe in a quiescent phase, we can only speculate about the duration over which the galaxy remains quenched; so we propose to call these objects (mini-)quenched, to indicate that it is likely that they will re-ignite again in its future, particularly for high-redshift or low-mass systems; but also leaving open the possibility that they continue to have a very low specific SFR (sSFR) over extended time-scales, or remain permanently quenched.

In order to unambiguously establish mini-quenching observationally, rejuvenating galaxies characterised by old stellar populations and UV-faintness while exhibiting strong nebular emission lines tracing recent star formation, have yet to be observed. Additionally, more galaxy spectra are needed to identify galaxies in different phases of their burstiness cycles: from "bursts" via "regular", and "lull" phases (see e.g. the post-starburst, nearly quiescent galaxy in \cite{Strait2023}), to mini-quenching and rejuvenating galaxies, to provide constraints on the physical processes shaping the burstiness of SFHs. Until such large data sets are obtained we propose to refer to these types of quiescent galaxy, which might re-ignite soon again in the future like the one presented in \cite{Looser+2023}, as (mini-)quenched, as argued above.

Large statistical samples are needed to characterize burstiness, (mini-)quenching events and the associated duty cycles as a function of different galaxy population properties, such as observed redshift, stellar mass \Mstar or distance from the main sequence (MS).\footnote{Where MS describes the positive scaling relation between \Mstar and SFR of the star-forming galaxy population \citep[][e.g.]{Sandles2022MNRAS.515.2951S}{}{}.}
%Within the fist year of observation with \jwst, these sample or just being assembled, 
Leaving the burstiness of high-redshift and low-mass galaxies - and the physics which shapes it - as one of the major unknowns in galaxy assembly and evolution to date; and one of the key science goals for \jwst.

In this paper, we present a first study of stellar populations in high redshift galaxies and present observational results on the burstiness of SFHs as a function of stellar mass and redshift; and further observational evidence for (mini-)quenching events at high redshift. This work is based on data acquired by our JWST Advanced Deep Extragalactic Survey (JADES, Eisenstein et al., in prep.) survey. JADES is a large JWST GTO program, formed out of a collaboration between the NIRSpec \citep{jakobsen+2022} and NIRCam \citep{Rieke2023} instrument science teams, combining both imaging and spectroscopy. The program is designed to present an unprecedented study of the physical properties of galaxies at high redshift, mostly focusing on targets beyond the Cosmic Noon.

The paper is structured as follows:
In section \ref{sec:Data} we present our JADES data, in particular the NIRSpec PRISM spectra, summarize the data processing and describe our full spectral fitting methodology based on \ppxf \citep{Cappellari2017,Cappellari2022}. In section \ref{sec:stellar_populations} we discuss our observational results about the stellar ages and stellar dust attenuation 
%and present stacked SFHs 
in different bins of characteristic galaxy properties. In section \ref{sec:Bursty_SFHs} we present our results on the burstiness of SFHs. In section \ref{sec:Discussion} we discuss our results. In section \ref{sec:Summary} we summarize the key findings of this paper.

Throughout this work, we assume a Chabrier \citep[][]{chabrier_galactic_2003} initial mass function (IMF) and a $\Lambda \textrm{CDM}$ cosmology with the following parameters: $H_0 = 70$ km $\textrm{s}^{-1}$/ Mpc, $ \Omega_{\textrm{M}} = 0.3$ and $\Omega_\Lambda = 0.7$.

\section{Data, data reduction and extraction of basic physical quantities} \label{sec:Data}
The NIRSpec \citep{jakobsen+2022} micro-shutter array \citep[MSA, ][]{Ferruit2022} spectra used in this work were obtained as part of our JADES GTO program (PI: N. Lützgendorf, ID:1210) observations in the Great Observatories Origins Deep Survey South field (GOODS-S, \citet{giavalisco2004}) between October 21--25, 2022.
These spectra constitute the HST-DEEP tier of the survey (hereafter: JADES/HST-DEEP) and were obtained using the disperser/filter configuration
{\sc prism/clear}, which covers the wavelength range between 0.6~µm and 5.3~µm and provides spectra with a nominal wavelength-dependent spectral resolution
of $R \ssim$30--330 \citep{jakobsen+2022}\footnote{For a sub-sample of targets, also higher-resolution grating spectra were obtained as a part of this program. However, we focus on the prism spectra in this work, which are more relevant for the stellar continuum.}.

The program observed a total of 253 galaxies over three dither pointings. For each target, three microshutters are simultaneously opened for exposure. Each dither pointing uses a different MSA configuration to place the spectra at different positions on the detector - to decrease the impact of detector gaps, mitigate detector artefacts and improve the signal-to-noise ratio (S/N) for high-priority targets, while increasing the density of observed targets. Within each individual dither, the exposure is three-pointing nodded along the slit. 
Each three-point nodding was integrated for 8403 seconds. The three-nod pattern has been repeated four times. Each target was observed in either one or multiple pointings, resulting in a total exposure time of up to 28 hours for the prism.

The flux-calibrated spectra were extracted using pipelines developed by the ESA NIRSpec Science Operations Team (SOT) and the NIRSpec GTO Team. A detailed description of the pipelines will be presented in a forthcoming NIRSpec/GTO collaboration paper (Carniani et al., in prep.). For a more detailed presentation of the JADES/HST-DEEP spectra and a discussion of the sample selection we refer to Bunker et al. (subm.). In this paper, we use all spectra for which a redshift could be established. For the galaxies with strong emission lines, we use the same redshifts as those presented in Bunker et al. (in prep.). Otherwise, we use redshifts inferred from visual inspection of the continuum and weaker lines, which are further refined by \ppxf, see below. 

We remark that with the effective slit width of 0.2 arcsec, the JADES
spectra suffer from wavelength-dependent aperture losses. This is
corrected assuming a point-source geometry, which leads to systematic
underestimate of the flux for the most extended, lowest-redshift sources.

\subsection{Full spectral fitting with \ppxf} \label{ppxf_methodology}

\begin{figure*}
\centering
\begin{subfigure} {1.0\columnwidth}
\includegraphics[width=\columnwidth]{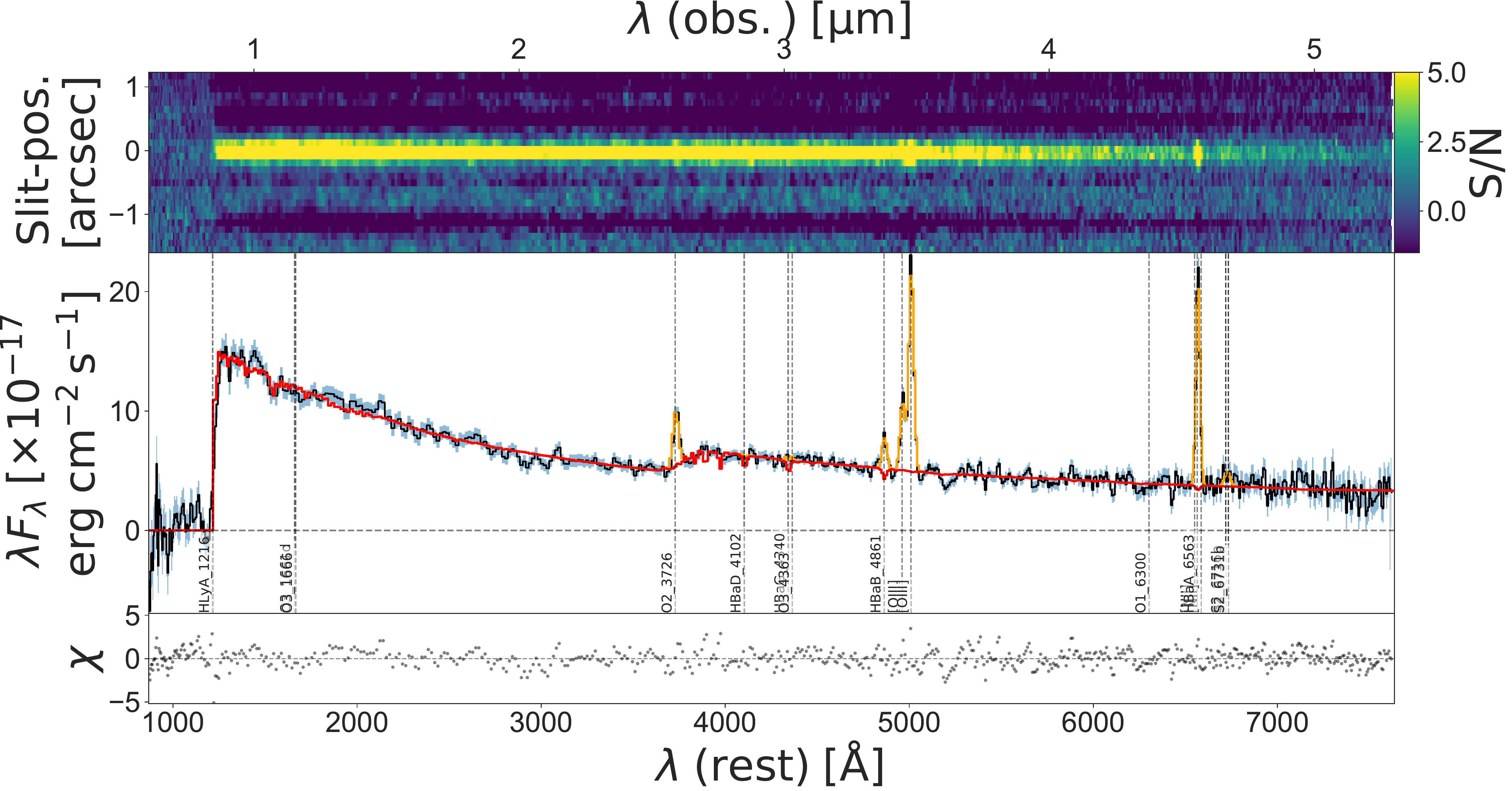}
\caption{} \label{fig:spectrum_10013618}
\end{subfigure} 
\hspace{\hdist mm}
\begin{subfigure}{1.0\columnwidth}
\includegraphics[width=\linewidth]{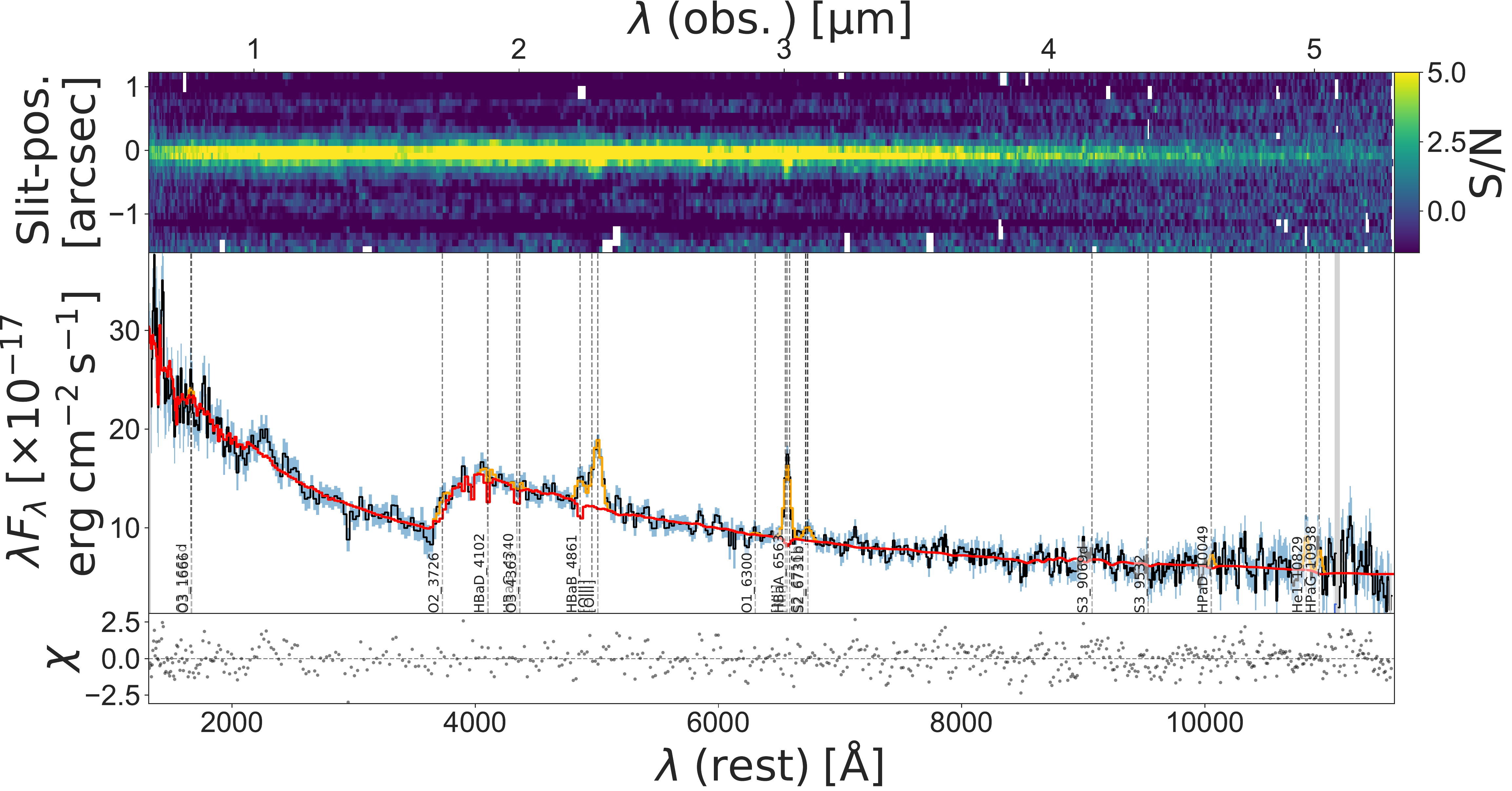}
\caption{} \label{fig:spectrum_005329}
\end{subfigure}
\caption{Left: Example of a JADES-DEEP spectrum of a star-bursting galaxy at redshift $z=5.9$. Right: Example of a JADES-DEEP spectrum of a weakly star-forming galaxy at redshift $z=3.6$. The noise is indicated by the blue shaded regions. The fitted \ppxf continuum is given in red, the nebular emission lines in yellow. The vertical dashed lines mark the rest-frame wavelengths of strong nebular emission lines. The upper panel shows the S/N of the combined 2D
    spectrum (note that the 1D spectrum is not extracted from the combined 2D spectrum). The bottom panel indicates the reduced residuals of the fit.} 
\label{fig:Example_fits_ppxf}
\end{figure*}
 
The R100 spectra are fitted with a methodology based on the $\chi^2$-minimization Penalized PiXel-Fitting code\footnote{\url{ https://pypi.org/project/ppxf/}; version 8.1.0} \ppxf \citep{Cappellari2017, Cappellari2022}, leveraging bootstrapping to infer key physical quantities
\citetext{Looser et al., subm.}. To fit the stellar continuum, a library of simple stellar-population (SSP) templates, coupled with a \cite{Calzetti2000} dust attenuation law (but without any additive or multiplicative polynomials), are fitted as a (non-negative) linear superposition to the continuum spectrum. The SSP library uses synthetic model atmospheres from the C3K library\citep{Conroy2019} with a resolution of R=10,000, adopting the MIST isochrones of \citet{Choi2016}, solar abundances and a Salpeter IMF. Throughout the paper, we change the IMF to Chabrier, using the formula of \cite{speagle+2014}. The synthetic SSP spectra span the 2D age-metallicity logarithmic grid from age$_{SSP}$ = $10^{6.0}$~yr to $10^{10.3}$~yr and [M/H] = -2.5 to 0.5. For each galaxy, we cut the age grid to be consistent with the age of the Universe at that redshift, plus a buffer of a single bin\footnote{These SSP templates are available from the author C. Conroy upon reasonable request.}. For a self-consistent treatment of nebular emission lines, we use Gaussians to fit them simultaneously with the stellar continuum -- the SSP models themselves do not
include nebular emission.

We describe below in detail the method based on the \ppxf algorithm applied to each JADES/HST DEEP spectrum in this work:
\begin{enumerate}
    \item First, the C3K templates are convolved to match the wavelength-dependant spectral resolution of the spectrum. Secondary to wavelength, the effective spectral resolution (R) depends on the degree
    of `slit filling', i.e. the ratio between the galaxy size and the 0.2-arcsec width of the
    micro shutters. This effect is estimated to be as large as a factor of two. To enable \ppxf to reproduce
    the data, we artificially increase R for all targets.
    \item Both the spectrum and the templates are re-normalized by the median flux per spectral pixel in the spectrum to avoid numerical issues, and to enable the use of regularization in \ppxf (`regul' keyword), allowing to penalize non-smooth weight distributions (see \citealp{Cappellari2017} for more details). 
    \item A first fit with regul=5 is used to obtain an initial estimate of the model and remove outliers using 4-$\sigma$ clipping.
    \item Then we perform a wild bootstrapping by perturbing the spectrum $S$ with the estimated noise spectrum from the data reduction $N$: $S^*(\lambda) = S(\lambda) \pm N(\lambda')$, where $N(\lambda')$ is randomly chosen from the noise spectrum within $\pm 50$ pixels for each spectral pixel $\lambda$. 
    \item We fit the perturbed spectrum $S^*(\lambda)$ again with \ppxf, again with regul=5.
    \item We repeat steps 4\&5 one hundred times.
    \item This method probes the sampling distribution of each individual SSP grid weight. The 100 bootstrapped grids of SSP weights are then averaged to recover a non-parametric SFH consistent with the intrinsic noise of the spectrum. 
\end{enumerate}

The output is a reconstructed assembly history of the target under consideration, as traced by this ``archaeological" approach using the observed remaining stellar populations as the ``fossil record" of the system. Due to the point-source assumption in the data reduction, our
spectra do not capture the absolute \Mstar and SFR of each galaxy,
leading to values that are systematically underestimated. In addition,
by modelling the spectra alone (i.e., without photometry), we neglect 
any light falling outside the MSA shutters; in particular, colour 
gradients and clumpy morphologies cannot be captured by our
approach. However, we remark that the impact of this effect is strongest 
where it is least relevant to our conclusions, i.e. in the lowest redshift 
bin and at the highest stellar masses.

The distinctive advantage of our approach is that, by fitting the observed spectra with a superposition of independent SSPs and gas templates, we are not imposing a particular parametric SFH or a single metallicity on the spectrum. In other words, our recovered SFHs are non-parametric and do not depend on any assumption about the underlying physics of galaxy evolution. Crucially, any recovered scaling relation cannot have been introduced by parametric assumptions about the shape of our fitted SFHs. 
As an example, the fitted light-weighted 2D grid of SSP-weights, and the conversion of these weights into a non-parametric SFH for the spectra presented in Fig.~\ref{fig:Example_fits_ppxf} is shown in Figs.~\ref{ppxf_SFH_example_2D}\&\ref{ppxf_SFH_example_1D}.

While this non-parametric astro-archaeological approach is extremely powerful, the analysis of its outputs has to be done with utmost care. This method has been tested on data from the local MaNGA \citep{Bundy2015} survey, and has been shown
to recover meaningful average star-formation and chemical-evolution histories \citetext{Looser et al., submitted}.

Focusing on the NIRSpec PRISM spectra, while the recovered ages, traced by the UV-slope, the Balmer break and the overall shape of the spectrum can be recovered reliably, the spectral resolution of the spectra may not be sufficient to reliably estimate the metallicity of the underlying galaxy populations. Due to the well known age-metallicity degeneracy, and known and unknown systematics, such as dust obscuration,  or flux calibration issues, we can trust returned SSP weights to different degrees. To assess the stability of the SSP weights, the bootstrapping methodology described above is highly instructive, as it returns a scatter distribution for each individual SSP weight, as well as for any quantity derived from the weight grid. The tests reveal that ages are reliable in conservative 1 dex bins in $\mathrm{log_{10}(Age [yr])}$, from $\mathrm{log_{10}(Age [yr])}$= 6.0 to 10.0. Hence, for this paper, we bin the SSP weight-grid, as presented in Fig.~\ref{ppxf_SFH_example_1D}, into four age bins, as presented below, see Fig.~\ref{fig:SFH_MW_bin_1}--\ref{fig:SFH_MW_bin_3}.

\subsection{Stellar mass}
To measure the stellar mass \Mstar for each galaxy, we sum the individual weights of the mass-weighted SSP grid fitted with \ppxf. To test the reliability of the \ppxf masses, we compare them to masses inferred from the \beagle code 
\citep{chevallard_beagle_2016} for the same data set, presented in \cite{Curti2023} and Chevallard et~al. (in~prep.). The \ppxf masses and the \beagle masses show a strong correlation with an RMS-scatter of ~0.2 dex, however we note an offset of ~0.2 dex. Even though the two codes infer different masses for some galaxies, the 
general agreement means that our choice of \Mstar does not drive the results presented in this paper. A comparison between the \ppxf and \beagle masses is presented in Appendix~\ref{comp_masses}.

\subsection{SFR from nebular Balmer lines}
\label{sec:mass_sfr}

To calculate the star-formation rate (SFR) from nebular emission lines, we follow a similar method as \citet{Curti2023}. We apply the calibration of \cite{kennicutt_star_2012}, using the attenuation-corrected \Halpha luminosity where available, i.e. at z$\lesssim$7, and 2.86$\times$\Hbeta otherwise. The dust-attenuation correction is based either on the Balmer decrement (\Halpha/\Hbeta=2.86 or \Hbeta/\Hgamma=0.47; where both lines are detected) adopting a \cite{Gordon+2003} dust correction; or on the E(B-V) from full spectral fitting, where only one Balmer line was detected.

Fig.~\ref{fig:Stellar-ages} shows the SFR-mass plane in three different redshift bins. The blue lines represent simple linear fits to the star forming mains sequence (MS) for this particular sample in the three redshift ranges. The fit likely overestimates the true MS due to complex selection effects. A more careful analysis of the MS, taking selection bias into account is beyond the scope of this work. Further, for the lowest, middle and highest redshift in each of the three redshifts bin, we plot the redshift-dependent quenching threshold $\mathrm{sSFR<0.2/t_{obs}}$, where $\mathrm{t_{obs}(z)}$ is the age of the Universe as a function of redshift \citep[e.g.][]{Gallazzi2014,pacifici+2016, carnall+2023b}. The quenching-threshold redshifts are $\mathrm{z=0,1,2}$ for the lowest redshift bin (colored in orange, green and red, respectively); $\mathrm{z=2,3.5,5}$ for the middle redshift bin; and $\mathrm{z=5,8,11}$ for the highest redshift bin.

\section{Results: Stellar populations} \label{sec:stellar_populations}
%\subsection{Distance from the star-forming Main Sequence}
In this section, we present the results of our non-parametric full spectral fitting with \ppxf on the inferred stellar population properties.

\subsection{Stellar ages as a function of \texorpdfstring{\Mstar}{Mstar}, \texorpdfstring{\DeltaMS}{Delta MS} and redshift z} \label{sec:stellar_ages}

\begin{figure*}
\centering
\includegraphics[width=\linewidth]{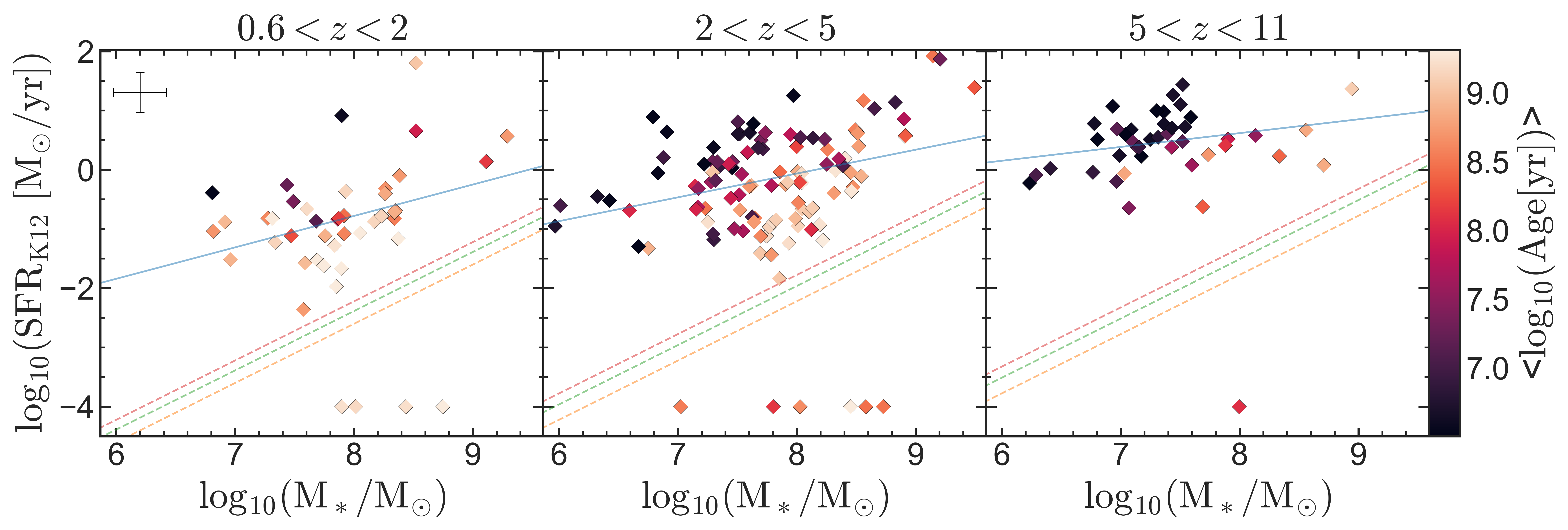}
\caption{SFR-mass plane color-coded by the average mass-weighted stellar ages measured by \ppxf in three different redshift bins. Each data-point represents a single galaxy. The blue lines represent a simple linear fit of the star-forming main sequence (MS) for this sample at that redshift bin. For reference, the three dotted lines indicate the quenched threshold for the lowest, middle and highest redshifts in each subplot, see main text. The error bar in the top-left corner represent the RMS-errors for \Mstar and SFR for the entire sample. Quiescent and (mini-)quenched galaxies, for which no SFR could be estimated due to non-detection of the relevant nebular emission lines, are plotted at the bottom of the three sub-figures. 
}
\label{fig:Stellar-ages}
\end{figure*}

In Fig.~\ref{fig:Stellar-ages}, individual galaxies are color-coded by their mass-weighted stellar ages. As one expects, with increasing redshift the average ages of the galaxies decrease. Further, we observe interesting trends with \Mstar and \DeltaMS: (a) High-mass galaxies tend to be overall older than low-mass galaxies in all redshift bins (in agreement with recent studies at Cosmic Noon, e.g. \citet{Carnall2019MNRAS.490..417C,Tacchella2022ApJ...926..134T,Ji+2022}), although there is considerable variation; (b) at fixed stellar mass, galaxies above the MS (as traced by nebular emission lines on 10 Myr time-scales) are younger than galaxies below the MS. This indicates that the \DeltaMS\ of a galaxy at the epoch of observation is to some extent correlated with its past formation history over longer time-scales. However, there is significant variation. The interesting question is whether this is due to measurement uncertainties or bursty SFHs, as will be discussed below.

\subsection{Stacked SFHs}

\begin{figure*}
   \centering
   \includegraphics[width=\linewidth]{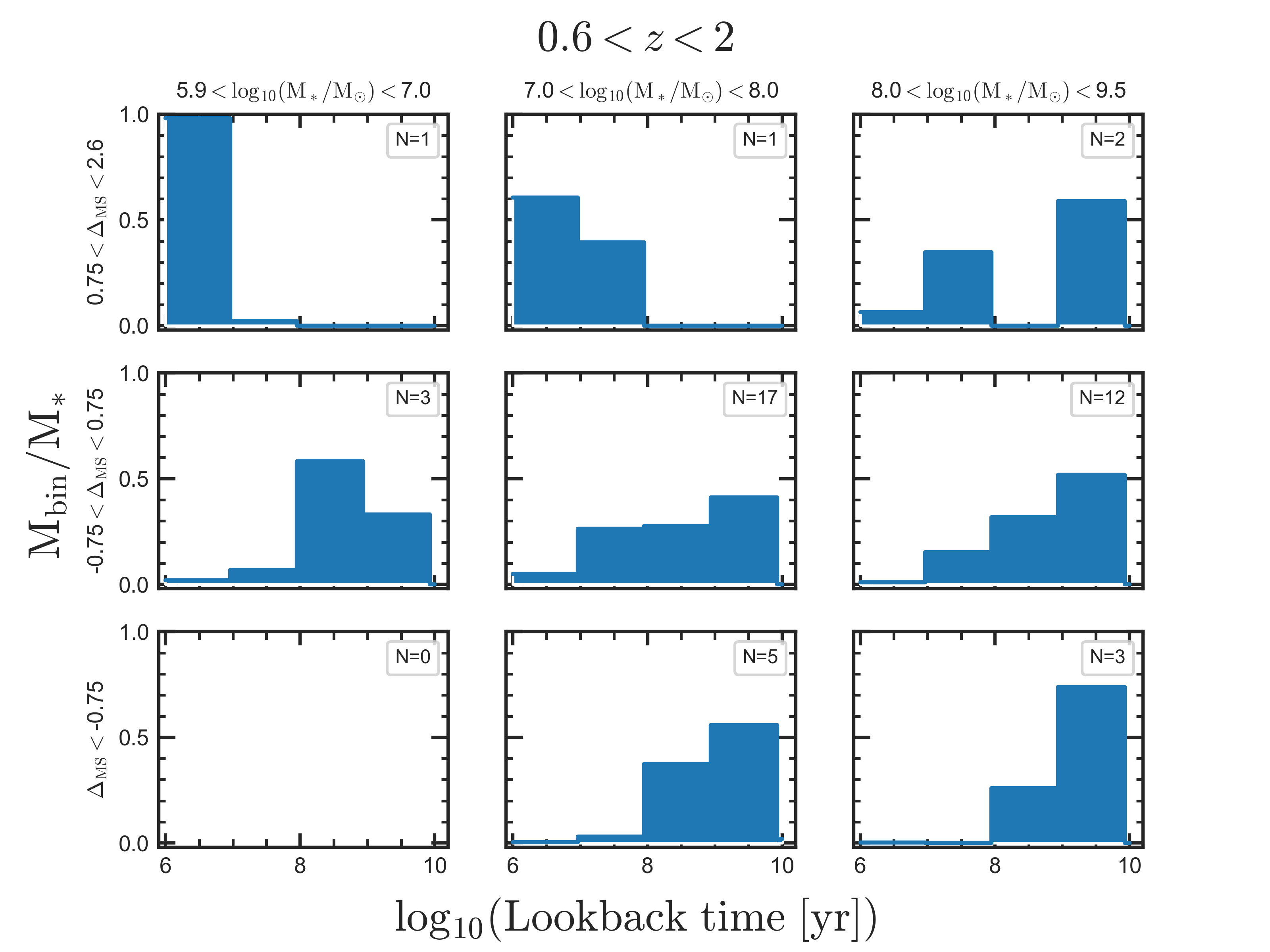}
    \caption{Mass-weighted stacks of "normalized" SFHs of galaxies in three bins of each, \Mstar and \DeltaMS, in the redshift range $\mathrm{0<z<2}$. The number in the legend of each sub-figure indicates the number of galaxies contributing to the stack. In each bin, the SSP-weights of each contributing galaxy are first normalized, and then averaged over all galaxies contributing to the bin. The underlying inferred SSP weight-grid  is collapsed into four age-bins, where each bin has a width of 1 dex, as indicated. The oldest bin (from 1 Gyr to 10 Gyr) is artificially extended for consistent plotting: For each individual galaxy SFH, only SSP templates consistent with the age of the Universe at the redshift of the target are included in the fitting. See section \ref{ppxf_methodology} for more details.}
    \label{fig:SFH_MW_bin_1}
\end{figure*}

\begin{figure*}
   \centering
   \includegraphics[width=\linewidth]{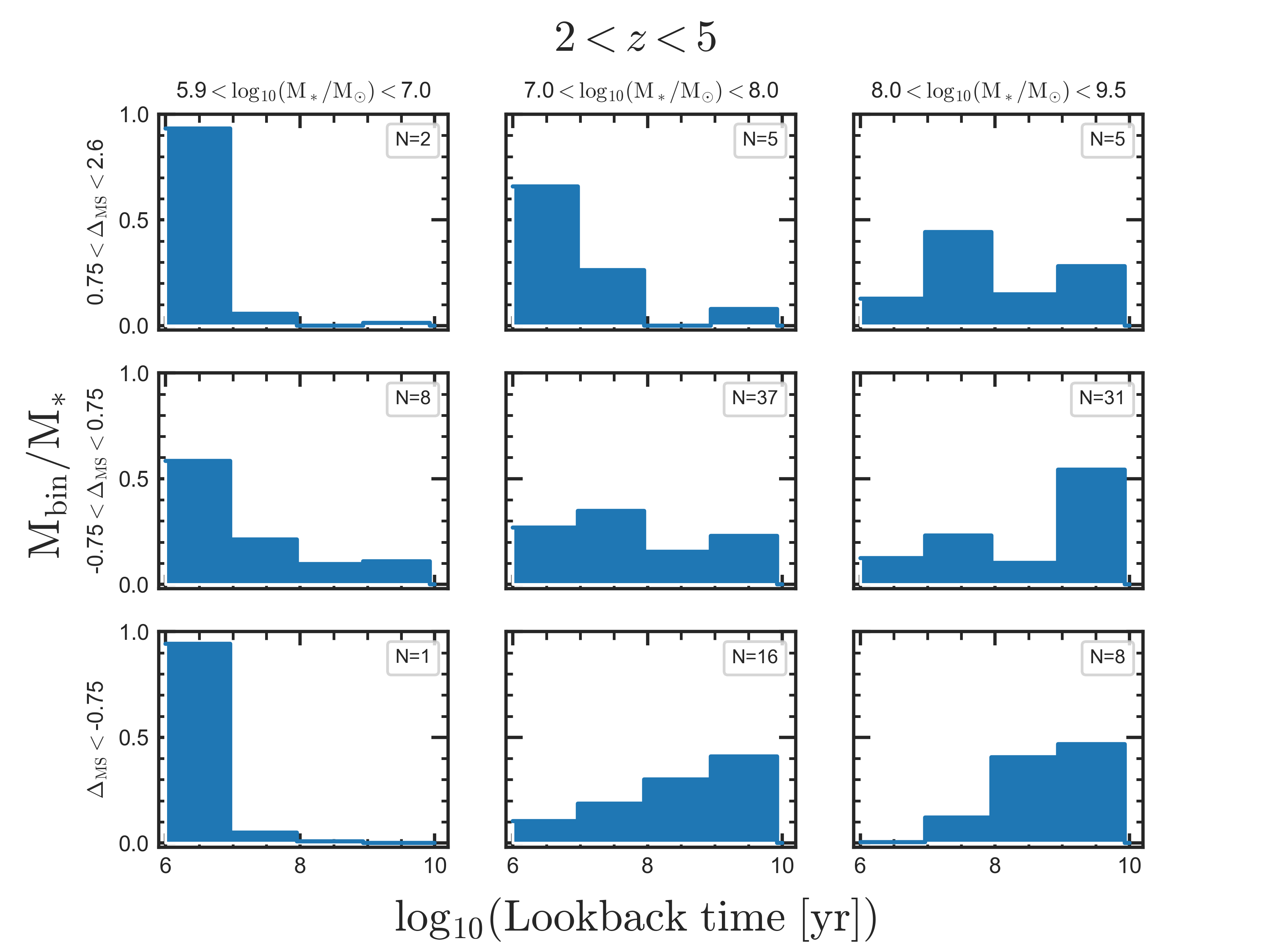}
    \caption{Mass-weighted stacks of "normalized" SFHs of galaxies in three \Mstar and \DeltaMS\ bins in the redshift bin $\mathrm{2<z<5}$. The number in the legend of each sub-figure indicates the number of galaxies contributing to each stack. See Fig.~\ref{fig:SFH_MW_bin_1} and section \ref{ppxf_methodology} for more details.}
    \label{fig:SFH_MW_bin_2}
\end{figure*}

\begin{figure*}
   \centering
   \includegraphics[width=\linewidth]{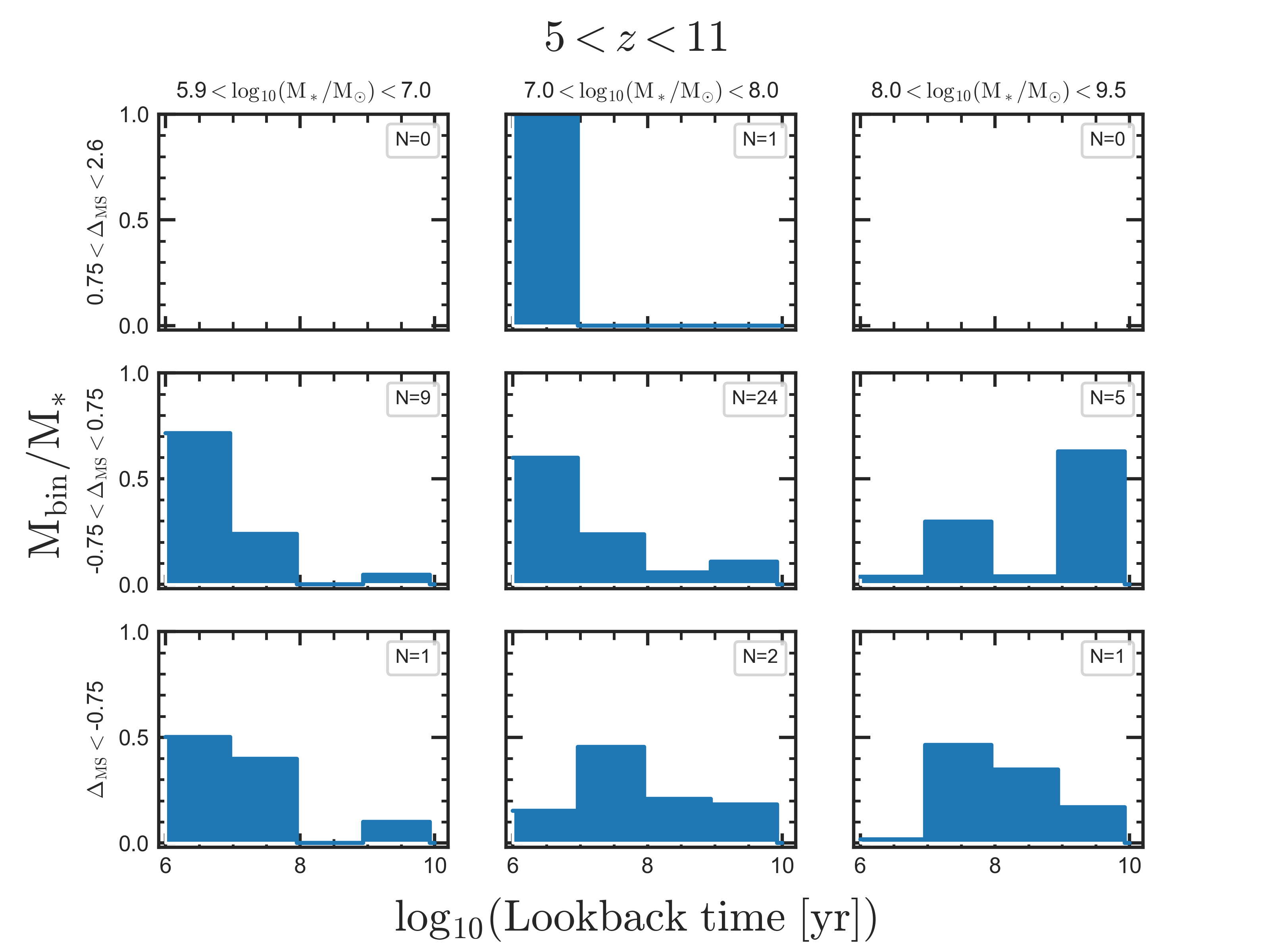}
    \caption{Mass-weighted stacks of individual "normalized" SFHs of galaxies in three \Mstar and \DeltaMS\ bins for galaxies with redshifts $\mathrm{z>5}$. The number in the legend of each sub-figure indicates the number of galaxies contributing to the stack. See Fig.~\ref{fig:SFH_MW_bin_1} and section \ref{ppxf_methodology} for more details.}
    \label{fig:SFH_MW_bin_3}
\end{figure*}

In Fig.~\ref{fig:SFH_MW_bin_1}-\ref{fig:SFH_MW_bin_3} the SFHs of individual galaxies are combined together to provide a composite view. The individual SFHs are stacked in three z, \Mstar and \DeltaMS\ bins, as indicated by the labels. We re-iterate that the individual \ppxf SSP-grid fits are non-parametric: \ppxf can freely choose the weighting of each individual SSP spectrum given, without any assumption on the functional form of the SFH. The stack in each bin are constructed as follows. First, for each galaxy, the SSP weights are normalized by the total sum of SSP-weights, i.e. we construct a "relative" weight-distribution of SSPs. These normalized weights are then averaged (i.e. each galaxy contributes equally) over all galaxies in a given  z--\Mstar--\DeltaMS\ bin. The inferred SSP-weight grids are then averaged over four log-age bins with width of 1 dex. This facilitates the identification of relative trends, and our bootstrapping tests indicate that the inferred grid weights cannot not be trusted on a finer age sampling. 

The stacked SFHs reveal interesting patterns in the SFHs of different galaxy populations in the sample as a function of z, \Mstar and \DeltaMS, similar to those found for more massive galaxies around the Cosmic Noon \citep{Ji+2023}. As one expects, low-redshift galaxies exhibit older populations, with more stellar mass formed at large look-back times. Conversely, high-redshift galaxies are mostly dominated by young stellar populations. Within each redshift bin, there are also interesting trends with \DeltaMS\ and \Mstar: Galaxies below the MS tend to have significantly more contribution of old SSPs than galaxies on the MS, while galaxies above the MS exhibit the highest mass fraction of young SSPs. Additionally, there is a weak trend with \Mstar: massive galaxies tend to be older than low-mass galaxies.

However, a very interesting finding is that even in the highest redshift bin there is a substantial contribution by evolved (old) stellar populations (formed more than one Gyr before the epoch of observation), indicating that the SFH analysis is revealing the imprint of the earliest episodes of star formation of these systems. However, we caution that inferring the oldest stellar populations in these system is quite difficult, as they contribute little to the stellar light, hence this finding should be confirmed with additional data.

\subsection{Dust attenuation as a function of \texorpdfstring{\Mstar}{Mstar}, \texorpdfstring{\DeltaMS}{Delta MS} and redshift z}

\begin{figure*}
\centering
\includegraphics[width=\linewidth]{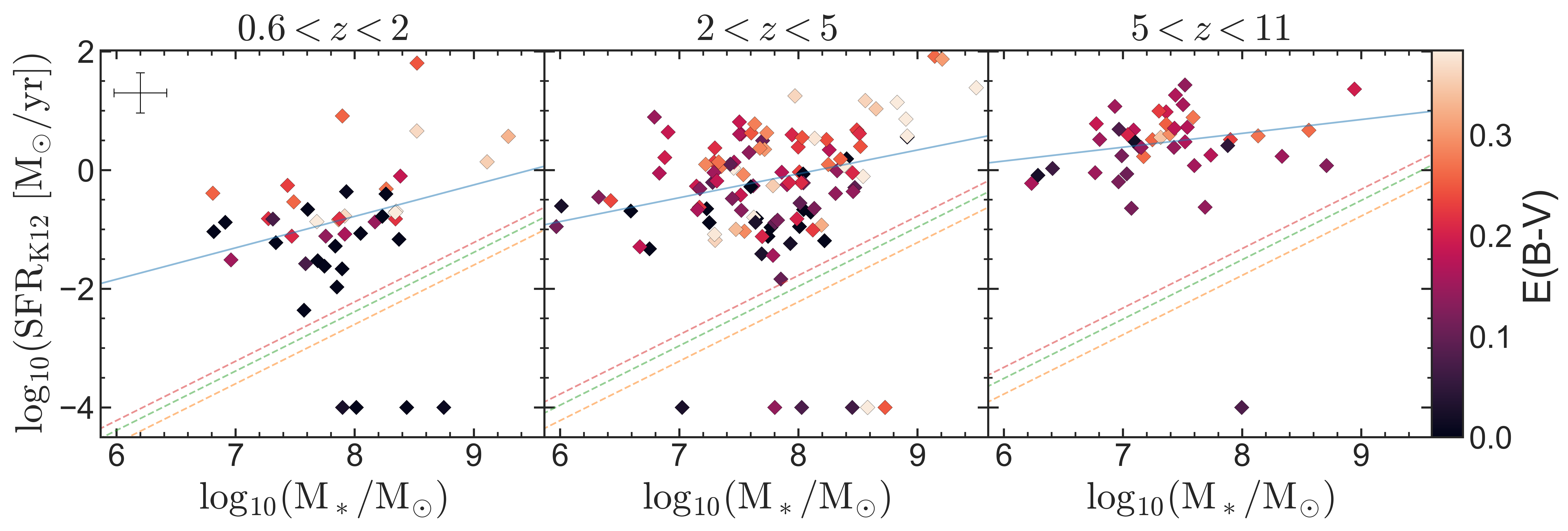}
\caption{SFR-mass plane color-coded by reddening E(B-V) of the stellar populations, inferred with \ppxf fitting of stellar populations convolved with a \citet{Calzetti1994} dust attenuation law with E(B-V) as a free parameter, in three different redshift bins. Each data-point represents a single galaxy. The quiescent galaxies are plotted at the bottom of the subplots. More details are given in Fig.~\ref{fig:Stellar-ages}.}
\label{fig:Stellar-reddening}
\end{figure*}

Fig.~\ref{fig:Stellar-reddening} shows the SFR-mass planes color-coded by E(B-V), tracing the amount of reddening of the stellar continuum caused by interstellar dust along the line of sight, as fitted with \ppxf. The galaxies are divided into the same three redshift bins as in section \ref{sec:stellar_ages}. We observe a clear trend with \DeltaMS: Spectra of quiescent galaxies and galaxies below the MS exhibit less dust attenuation, while particularly star-bursting galaxies exhibit significant reddening (see also Sandles et al., in prep.). Additionally, there is a trend with \Mstar: massive galaxies tend to be dustier than low-mass galaxies. Finally, the dust reddening declines at the highest redshift (at a given stellar mass and SFR), although most galaxies in the highest redshift bins still exhibit the presence of some, but moderate amounts of dust, especially in the more star forming and massive systems. In the two lower-redshift bins, most galaxies below the MS exhibit no dust at all, while particularly high-mass, star-bursting galaxies above the MS show significant dust reddening of the stellar continuum. Overall, we recover the mass dependence observed in the local Universe and at low redshift \citep[e.g., ][]{Pannella2009,Whitaker2017,McLure2018,Shapley2022,Maheson23}, %maheson+2023
but also show a clear dependence on the SFH and/or redshift.

\subsection{SFR from stellar populations}\label{sec:SFRs_ppxf}

\begin{figure*} 
\centering
\begin{subfigure}{.90\columnwidth}
\includegraphics[width=\columnwidth]{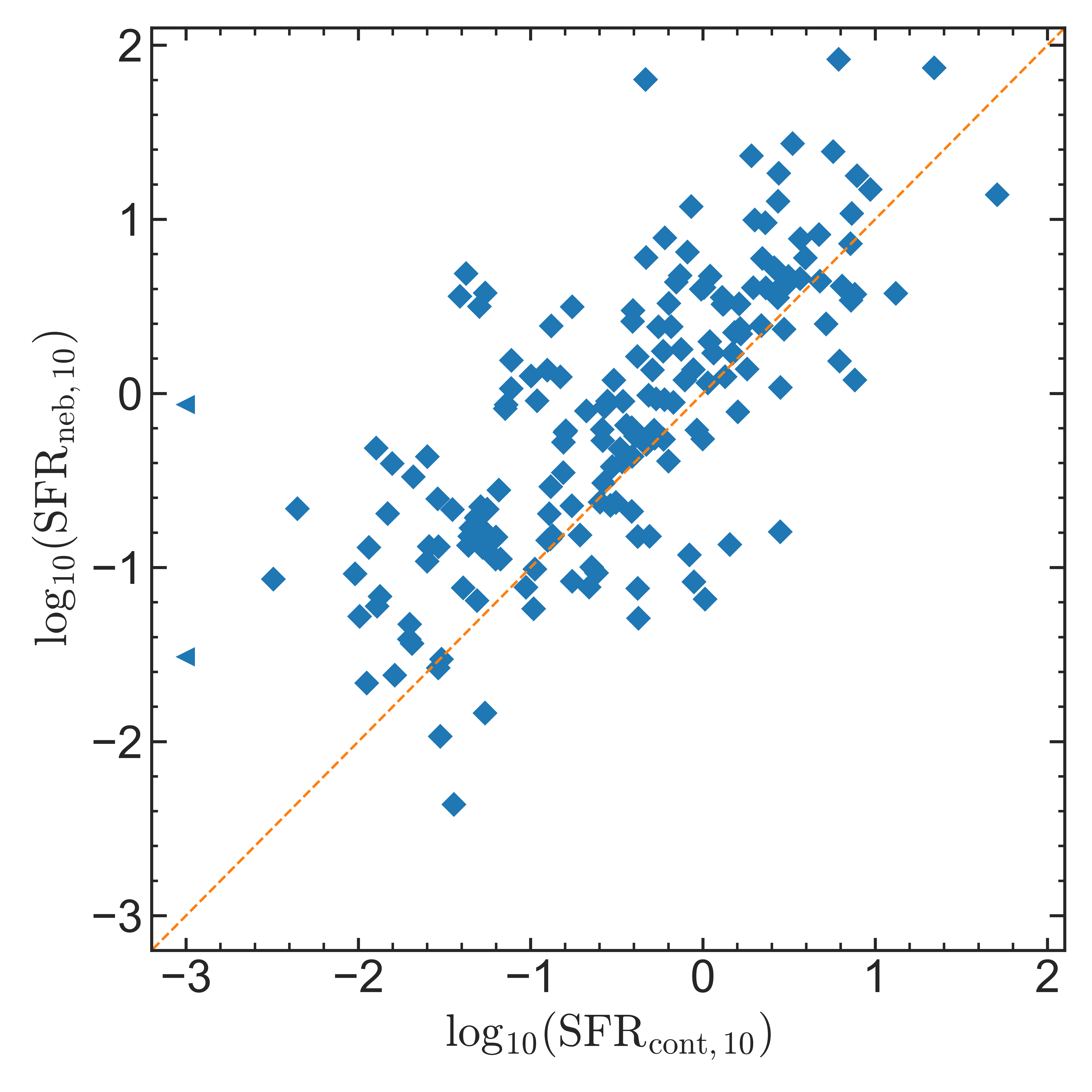}
%\caption{SFR estimated from stellar population fitting with \ppxf averaged over 100 Myr.} 
\end{subfigure}
\hspace{\hdist mm}
\begin{subfigure}{.9\columnwidth}
\includegraphics[width=\columnwidth]
{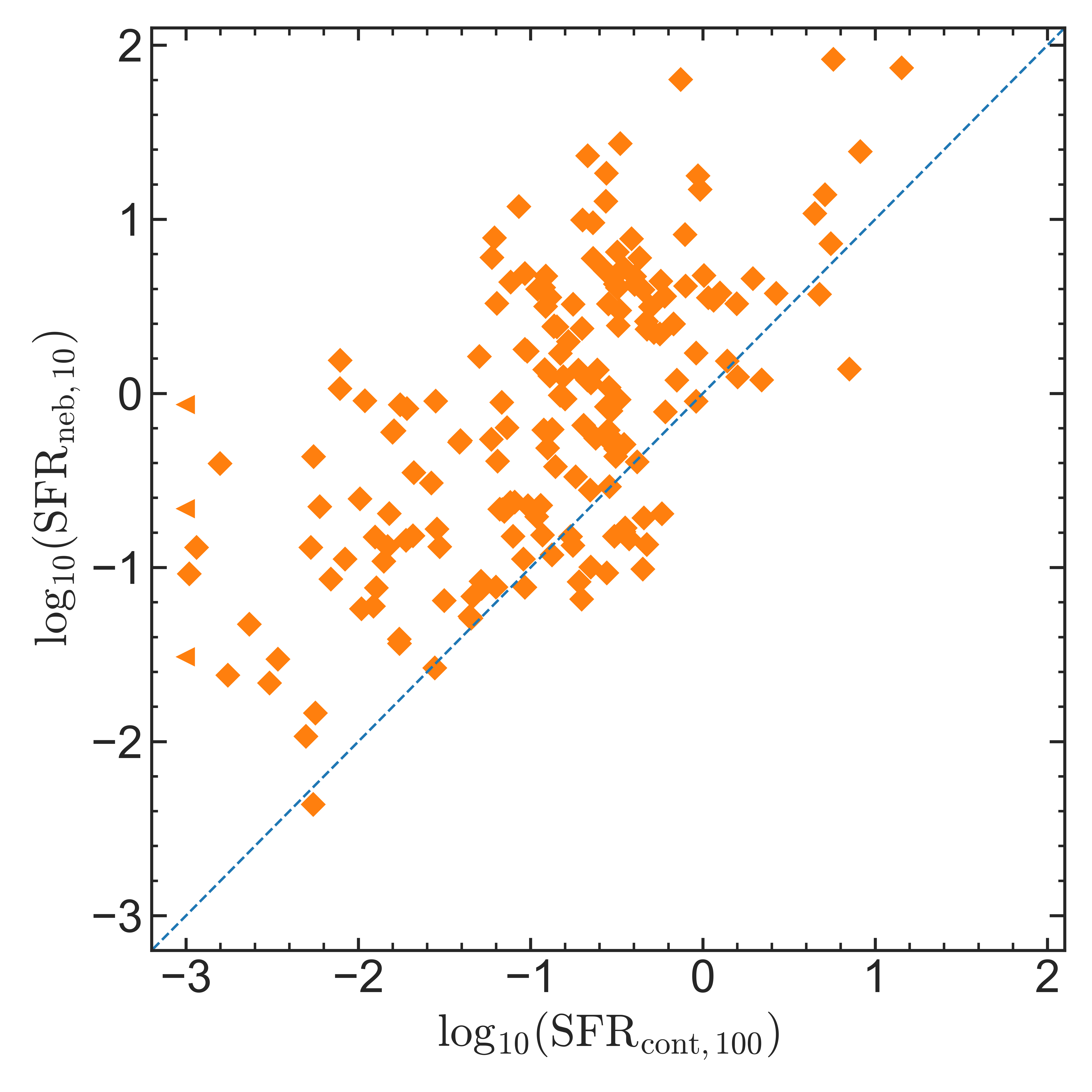}
%\caption{SFR estimated from stellar population fitting with \ppxf averaged over 10 Myr.} 
\end{subfigure}
\centering
\caption{SFR measured from nebular emission lines, tracing SFR over $\sim$10 Myr time-scales, using the K12 relation (y-axis, see section \ref{sec:mass_sfr} for more details) versus SFR measured from the non-parametric stellar population fitting of the continuum with \ppxf (x-axis). Left: SFR estimated from stellar population fitting with \ppxf averaged over the last 10 Myr before observations. The RMS-scatter between the two measurements is 0.3 dex and we note an offset of 0.2 dex, see text. Right: SFR estimated from stellar population fitting with \ppxf averaged over 100 Myr before observations. The interesting question is how much of the scatter and offset between these two quantities stems from noise and measurement uncertainty, and how much comes from physical variability in SFR plus selection bias (see \citet{Sun+2023} for a discussion). As we will argue below, a significant portion of the scatter is of physical origin, suggesting bursty SFHs.} \label{SFR_ppxf_10_100_comp}
\captionsetup{width=0.5\columnwidth} 
\end{figure*} 

In Fig.~\ref{SFR_ppxf_10_100_comp} we present results on the inference of $\mathrm{SFR_{cont}}$ over different time-scales directly from the stellar populations and compare this to SFR estimated from nebular emission lines (\sfrnebten), see section \ref{sec:mass_sfr}, where $\mathrm{SFR_{cont}}$ is derived via the fitting of the stellar continuum with \ppxf; and averaged over 10 Myr (\sfrcten, left) and 100 Myr (\sfrchund, right), respectively. 

We find excellent agreement between \sfrcten traced by the stellar continuum, and \sfrnebten traced by the optical Balmer lines. The small difference in normalisation might be partially explained by the assumption of solar metallicity in K12, which overestimates the SFR in metal-poor systems. A comparison between K12 and metallicity calibrated SFR, see e.g. \cite{shapley2023}, will be presented in a forthcoming work; for this study, we focus
on the mean trend and its scatter, therefore we ignore the systematic offset.

The right panel of Fig.~\ref{SFR_ppxf_10_100_comp} shows a strong correlation between \sfrchund and \sfrnebten, although there is more scatter, and we observe a deviation in the normalization.

The interesting question is whether the scatter is due to physical variation in SFR, i.e. bursty SFHs, or due to measurement uncertainties. Moreover, the difference in overall normalization may be due to selection bias, i.e. we preferentially observe galaxies in "star-bursting" phases, as they are much brighter, both in nebular emission line luminosities and UV continuum luminosity. We will present observational evidence that a significant part of the scatter is physical, indicating bursty SFHs particularly at high-redshift and in low-mass systems, in the next section.

\section{Results: Observational evidence for bursty SFHs}\label{sec:Bursty_SFHs}
In this section, we present observational results indicating bursty SFHs in high-redshift and low-mass systems. Specifically, in addition to the star-bursting galaxy in Fig.~\ref{fig:spectrum_10013618}, we show examples of galaxies in "regular" phases, i.e. galaxies which formed stars at roughly the same rate over the last 10 Myr as over the last 100 Myr; galaxies in "lull" phases, i.e. galaxies which formed significantly fewer stars over the last 10 Myr relative to the last 100 Myr, but with detected Balmer lines associated with recent star-formation activity and above the quenched-threshold ($\mathrm{sSFR}\gtsim \mathrm{0.2/t_{obs}}$). More broadly, we present an analysis of burstiness as a function of redshift $z$ and in the SFR-mass plane. Furthermore, we present the discovery of an additional, low-mass (mini-)quenched galaxy in a quiescent phase at high redshift. Finally, we also discuss observational biases.

\subsection{Examples of galaxies in "lull" or "regular" phases at high-z} \label{Lull_normal_examples}

\begin{figure*}
\centering
\begin{subfigure}{1.0\columnwidth}
\includegraphics[width=\columnwidth]{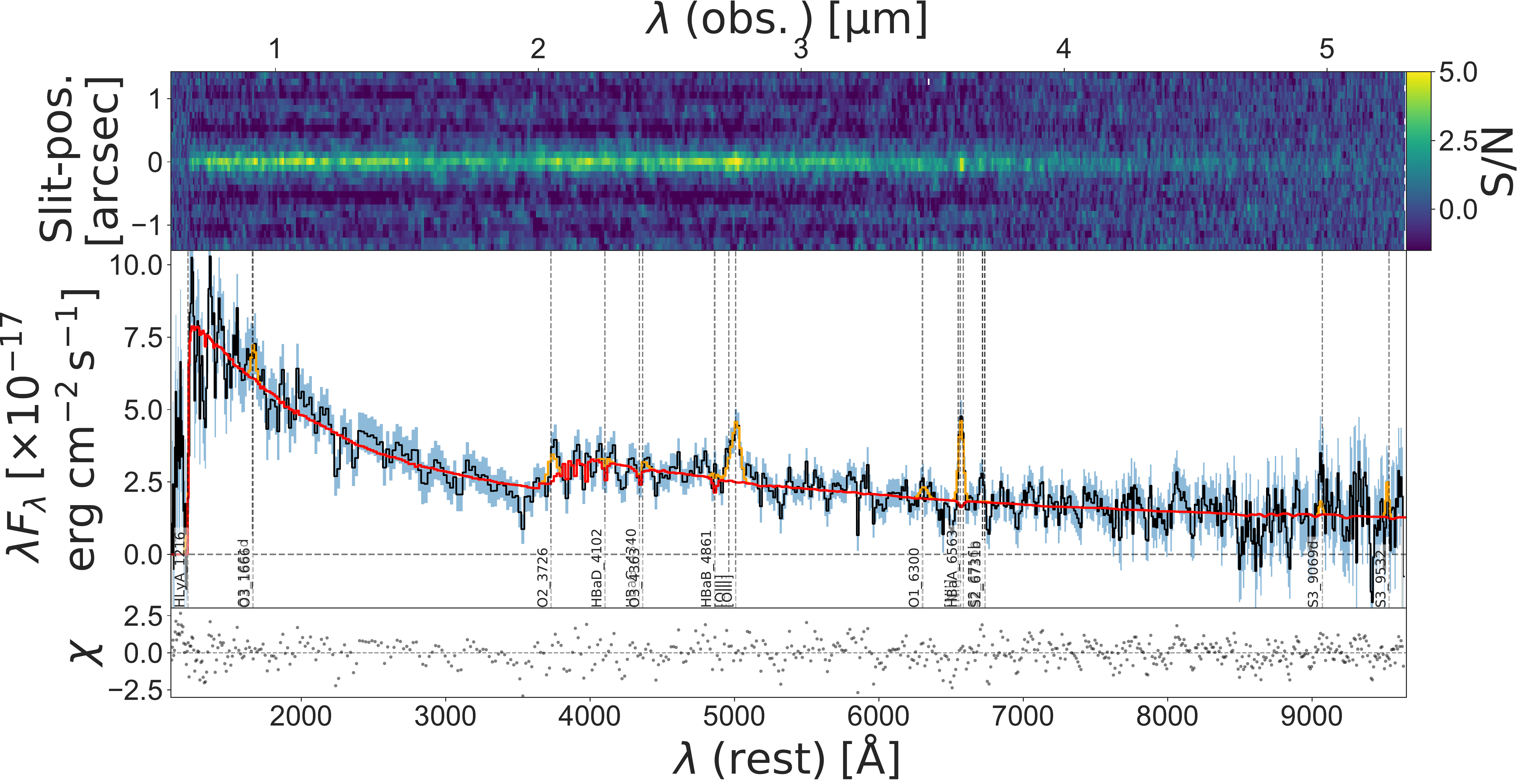}
\caption{Example of a galaxy in a "lull" phase at redshift $z = 4.5$. The spectrum is blue with a quite steep UV slope, but low EW in \Halpha, and [OIII] and exhibits a quite strong Balmer break.}  \label{lull_example}
\end{subfigure}
\hspace{\hdist mm}
\begin{subfigure}{1.0\columnwidth}
\includegraphics[width=\columnwidth]{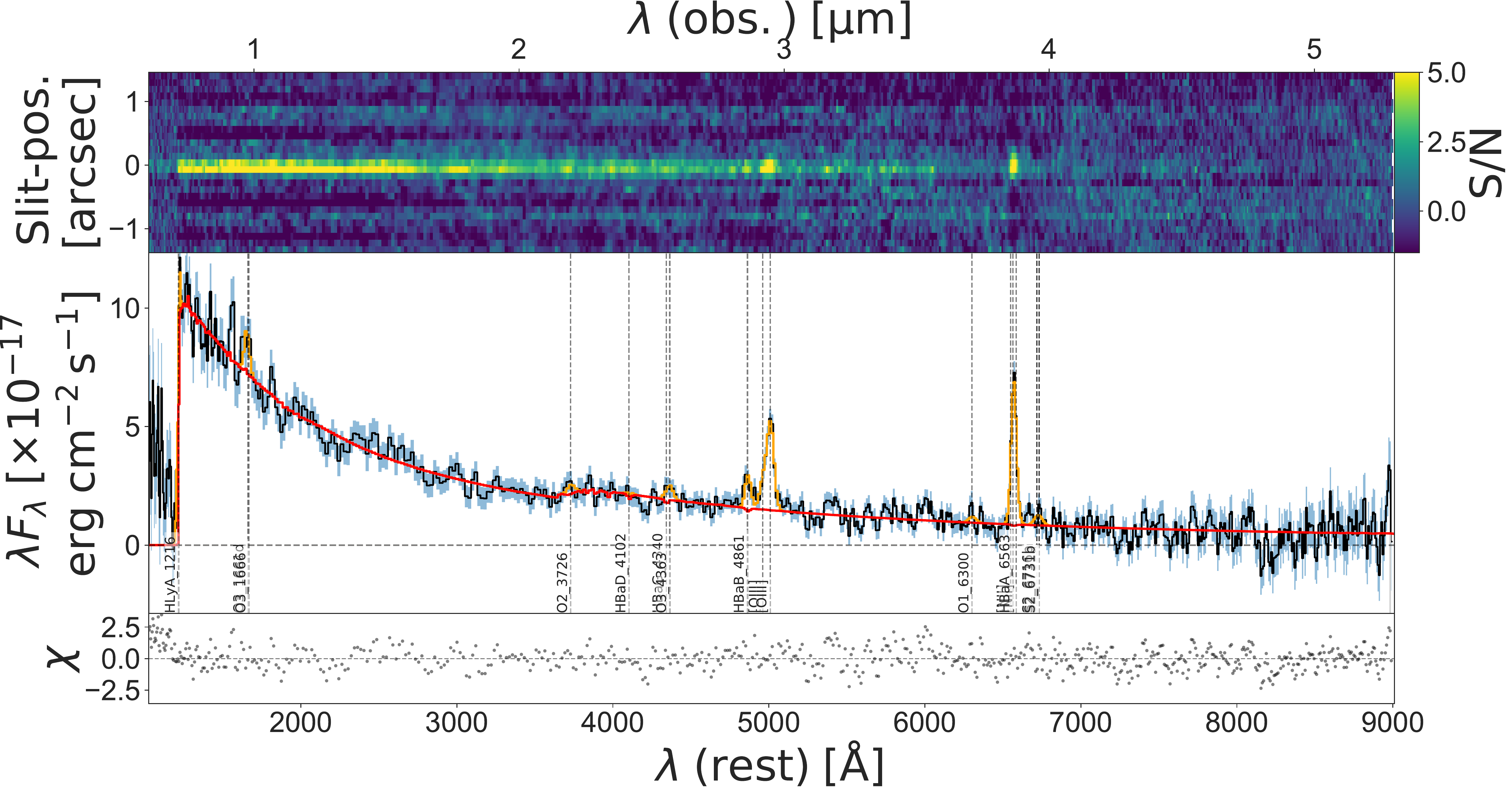}
\caption{Example of a galaxy in a "regular" phase at redshift $z = 4.9$. The spectrum is blue with a steep UV slope, but quite low EW in \Halpha, \Hbeta, and [OIII], and exhibits a marginal Balmer break. } \label{normal_example}
\end{subfigure}
\centering
\caption{Observational evidence for bursty SFHs in JADES/HST-DEEP. Left: A galaxy in a "lull" phase. Right: A galaxy in a "regular" phase. Both galaxies exhibit low EW in nebular emission lines.
}
\captionsetup{width=0.5\columnwidth}
\label{fig:Bursty_SFH_examples}
\end{figure*}

Fig.~\ref{fig:Bursty_SFH_examples} shows two examples of observed JADES/HST-DEEP galaxies in a "lull" and a "regular" phase, respectively, in the intermediate-redshift bin. Although both spectra are blue with a steep UV slope, indicating strong star formation over the past $\sim$100 Myr, the nebular emission lines are low-luminosity, indicating lower star-formation activity over the last $\sim$10 Myr before the epoch of observation for the galaxy in Fig.~\ref{lull_example}, and a regular star-formation activity over the last 10 Myr in the galaxy in Fig.\ref{normal_example}. The regular galaxy shows low EW emission in \Hbeta, [OIII] and \Halpha, for these redshifts. The lulling galaxy show even lower EW, particularly in \Halpha, and no detection of \Hbeta. The first one shows only a very weak Balmer break whereas in the latter an already quite strong Balmer break is emerging. This further supports the interpretation of these galaxies as "in a regular" phase and "in a lull" phase.

\subsection{Burstiness as a function of redshift \texorpdfstring{$z$}{z} and \texorpdfstring{\Mstar}{Mstar}}

\begin{figure*}
\centering
\includegraphics[width=\linewidth]{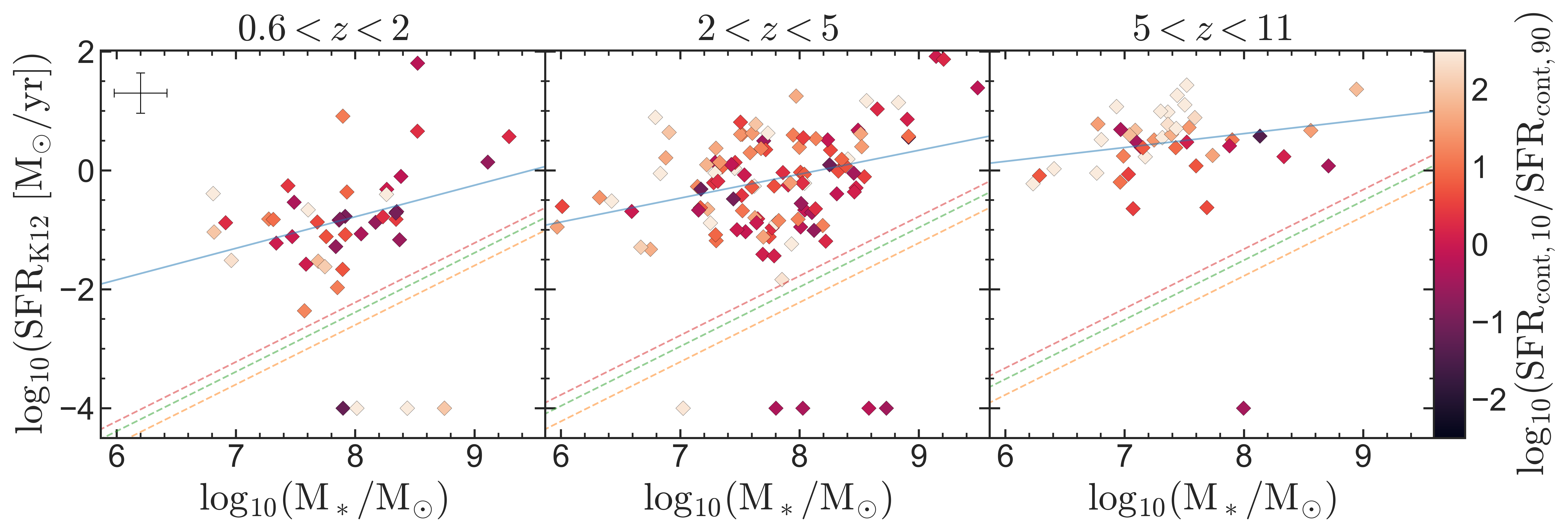}
\caption{Observational evidence for bursty SFHs: SFR-mass plane color-coded by the ratio of the SFR over the last 10 Myr (\sfrcten), and 10 Myr to 100 Myr (\sfrcninety) before the epoch of observation. Both tracers are inferred from non-parametric SSP fitting of the stellar continuum with \ppxf. Each data-point represents a single galaxy. The quiescent galaxies are plotted at the bottom of the subplots. More details are given in Fig.~\ref{fig:Stellar-ages}.}
\label{SFR_mass_plane_burstiness}
\end{figure*}

Fig.~\ref{SFR_mass_plane_burstiness} shows the ($\sfrnebten$)-mass planes color-coded by the ratio \sfrcten/\sfrcninety, i.e. the ratio between the SFR averaged over the last 10~Myr and the SFR averaged between 10 to 100 Myr. The data is divided in the three redshift bins, as indicated. Both, \sfrcten and \sfrcninety are inferred with stellar population fitting with \ppxf. The reason for using \sfrcninety is that it is estimated from a distinct set of weights in the SSP-grid, i.e. we avoid correlation by construction.

The ratio between the two \ppxf SFR tracers over different time-scales indicates whether a single galaxy is in a "burst", a "regular" or a "lull" phase at the epoch of observation. Studying the variation of this ratio between galaxies with otherwise similar properties provides important evidence for the burstiness of this galaxy population.

%\subsection{Burstiness as traced by nebular lines and stellar populations}

\begin{figure*}
   \centering
\begin{subfigure}{1.\columnwidth}
\includegraphics[width=1.\columnwidth]{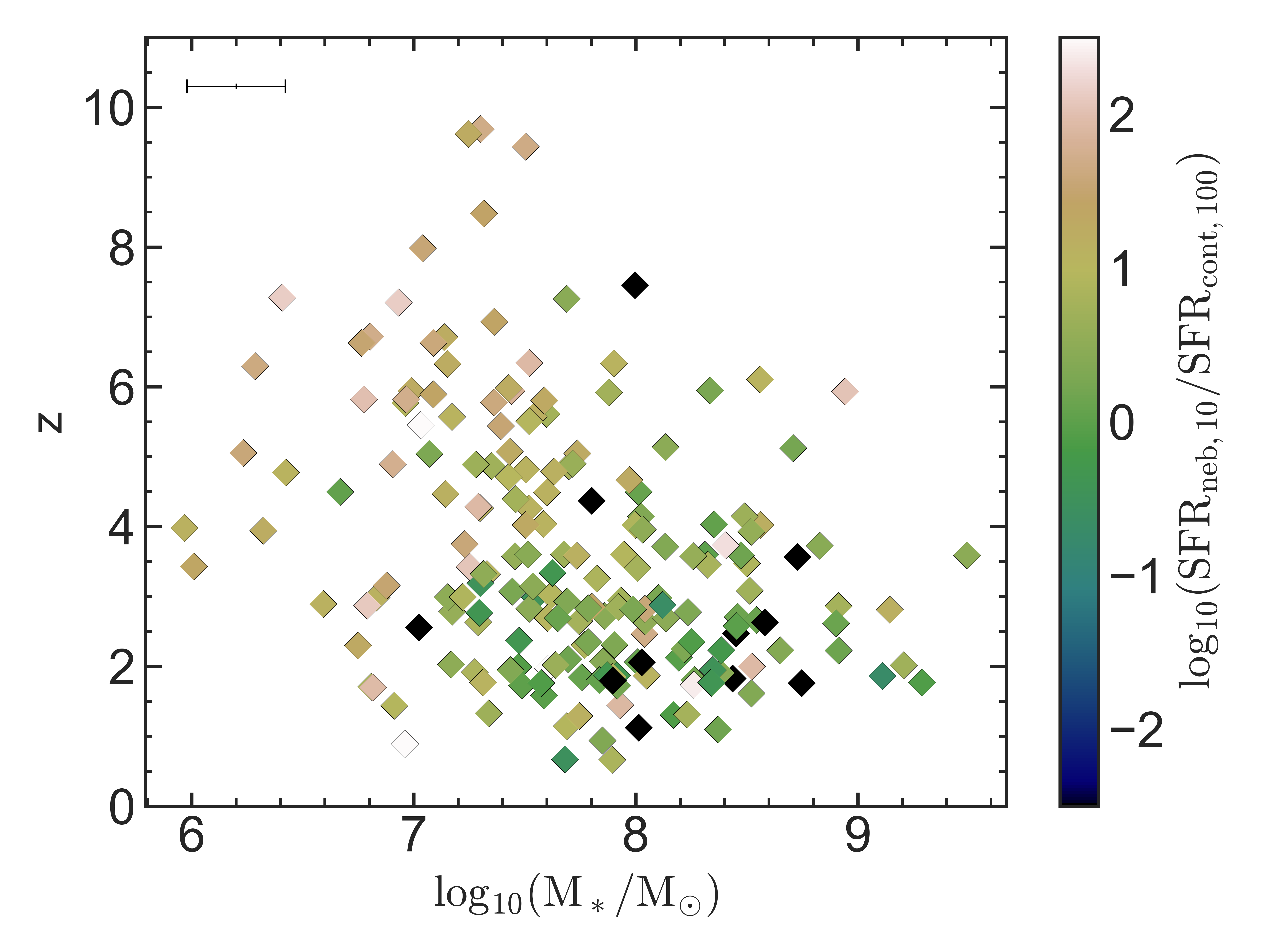}
\end{subfigure}
\begin{subfigure}{1.\columnwidth}
\includegraphics[width=1.\columnwidth]{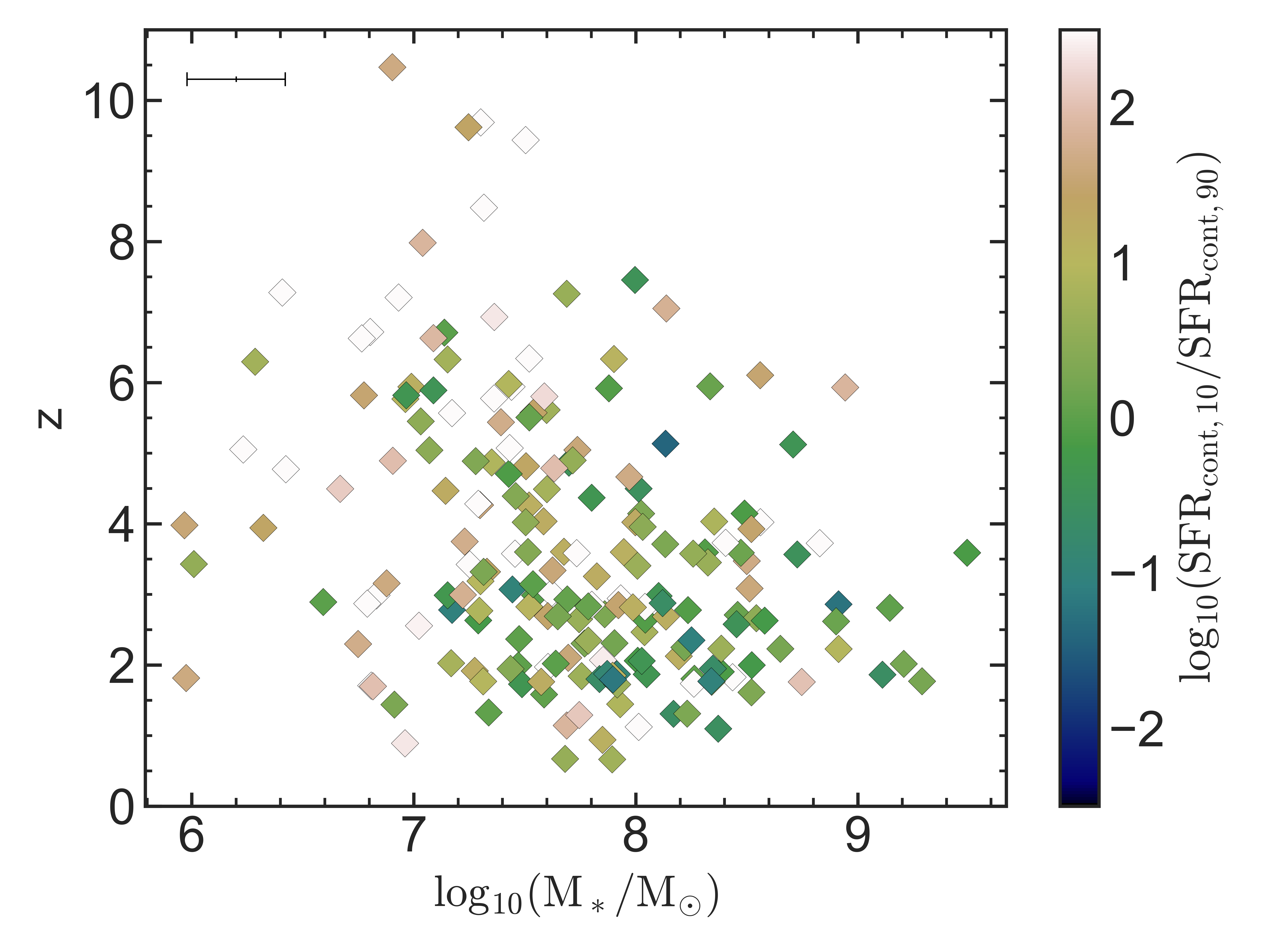}
\end{subfigure}
    \caption{Stellar mass and redshift dependence of burstiness in galaxy SFHs as inferred from the JADES/HST-DEEP sample. Left: The color-coding indicates the average star-formation over the last $\sim$10 Myr, as traced by Balmer emission lines, relative to the average SFR over the last 100 Myr, as traced by the stellar populations. Right: Relative SFRs inferred from the stellar populations; averaged over the last 10 Myr, and 10 Myr to 100 Myr before the epoch of observation (indicated as \sfrcten and \sfrcninety respectively). Green color-coded galaxies are in a regular state at the epoch of observation, white/light galaxies are in a burst, dark galaxies are in a lull, and black galaxies are (mini-)quenched.
    } 
    \label{fig:Bursty_SFH}
\end{figure*}

A complementary perspective, and important validation, of our observational evidence for the variation of star-formation burstiness with redshift z and \Mstar is presented in Fig.~\ref{fig:Bursty_SFH}. Galaxies in "regular" phases are indicated by green colors; galaxy in burst phases by light colors; galaxies in lulls in dark colors, and (mini-)quenched or permanently quenched galaxies in black. Here we investigate burstiness by looking at the ratio of $\sfrnebten$ and \sfrchund (see left plot). The right plot shows \sfrcten/\sfrcninety, as in Fig.~\ref{SFR_mass_plane_burstiness}. 
These results are consistent with the results presented above: it is evident that the low-mass and high-redshift galaxy populations are burstier, while the high-mass and low-redshift populations exhibit more galaxies in "regular" phases. As described above, in the low-mass and high-redshift populations, we predominantly observe galaxies in a burst phase, which is likely an observation-bias \citep{Sun+2023}. Crucially, the two plots show strong evidence for bursty SFHs based on two different methods: Left: \sfrnebten/\sfrchund traces burstiness using independent information from the nebular Balmer lines and the stellar continuum (similar to estimating the average SFR over 100 Myr time-scales directly from the UV-luminosity); Right: \sfrcten/\sfrcninety traces bursty SFHs self-consistently from the information about the stellar continuum alone.
However, in order to confirm the "bursty SFH" scenario, galaxies in mini-quenched phases have to be observed. Observational evidence and a discussion for this is presented in the next subsection.

\subsection{Discovery of another (mini-)quenched galaxy at high redshift}

\begin{figure*}
   \centering
   \includegraphics[width=\linewidth]{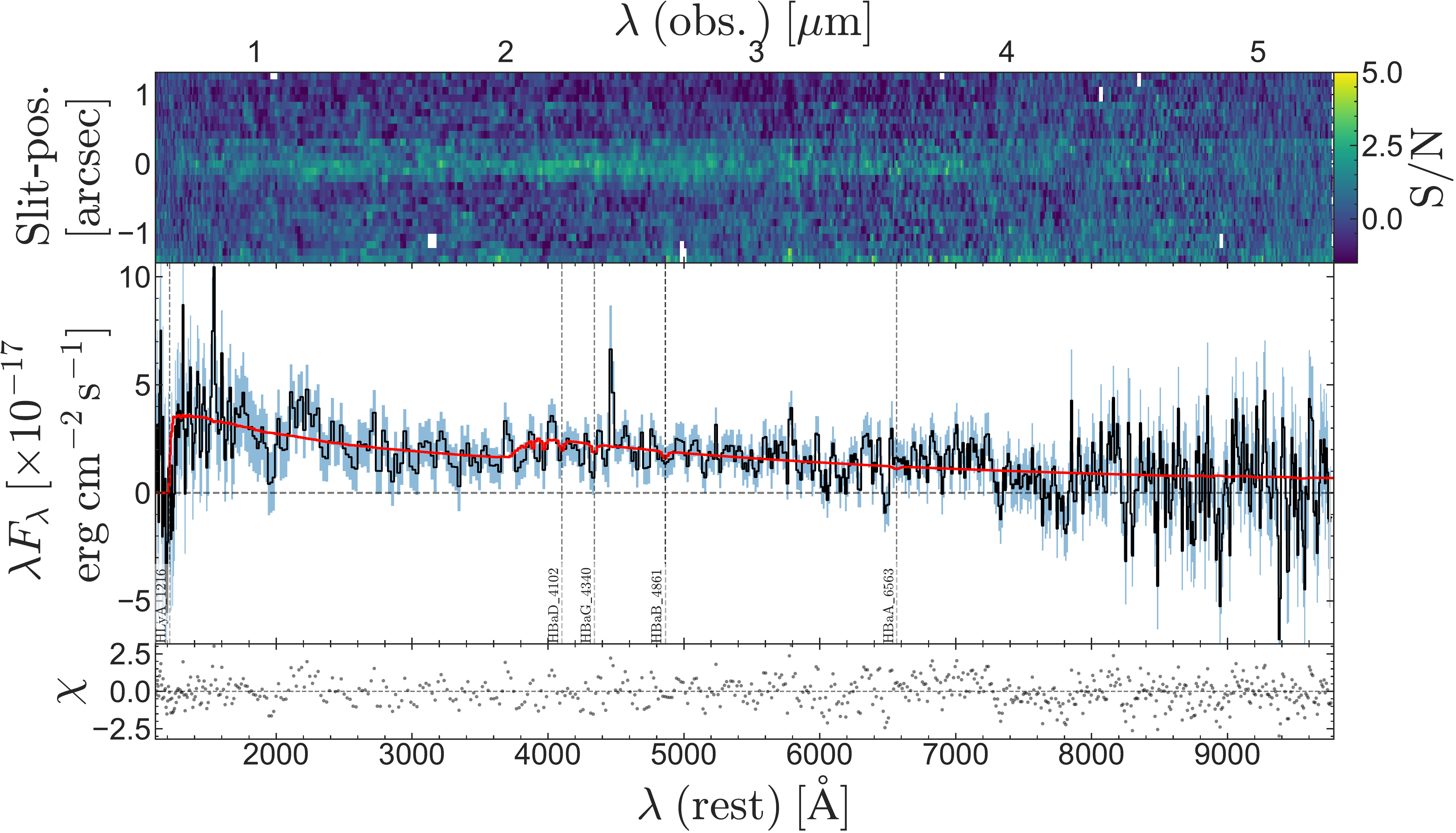}
    \caption{More spectroscopic evidence for (mini-)quenching at high-redshifts: The \LymanAlpha drop and the Balmer break clearly establish the redshift at $z=4.4$. The absence of emission lines suggests that this galaxy is in a quiescent phase (which is likely only temporarily) and has not formed a  significant amount of stars over the last 10 Myr before the epoch of observation.} 
    \label{fig:Mini-quenched}
\end{figure*}

In Fig.~\ref{fig:Mini-quenched} we present another high-redshift non-, or only weakly, star-forming galaxy with clearly determined redshift of $\rm z=4.4$, which is in addition to the post-starburst galaxy presented in \cite{Strait2023} and the fully (mini-)quenched galaxy \targetid in \cite{Looser+2023}. As in the spectra of those two, we observe a clear \LymanAlpha drop and a weak Balmer break. The stellar continuum of the spectrum is similar to the \citet{Strait2023} object at $z=5.2$, at slightly lower redshift and with a stellar mass of $\Mstar=10^{7.8}\ \M_{\odot}$. However, as in \targetid, there is no evidence for ongoing star formation on 10 Myr time-scales, as traced by nebular Balmer emission lines. It should be noted that the stellar mass inferred for \targetid by our \ppxf methodology is $\Mstar=10^{8.2}\ \M_{\odot}$, i.e. $\sim$ 0.5 dex lower than the masses inferred by the other three codes in \cite{Looser+2023}.

The finding of this additional (mini-)quenched galaxy, supports the scenario in which extreme burstiness can even lead to complete suppression of star formation, at least for short periods of a few 10 Myr.

\section{Discussion} \label{sec:Discussion}

% In this section, we discuss the results on high-redshift stellar populations presented in section \ref{sec:stellar_populations} and the results on bursty SFHs presented in section \ref{sec:Bursty_SFHs} inferred from the JADES/HST-DEEP sample.

\subsection{Short-timescale SFR from stellar continuum}

Our approach to measure SFHs based on \ppxf is substantially different compared 
to Bayesian inference methods \citetext{to name only a few, \beagle, \citealp{chevallard_beagle_2016}; \prospector, \citealp{johnson+2021},
\bagpipes, \citealp{Carnall2018}; \textsc{fado}, \citealp{Gomes2017};
\textsc{cigale}, \citealp{Noll2009}; and \textsc{prospect}, \citealp{Robotham2020}}. To some extent, all these softwares aim to
reduce the number of degrees of freedom that determine a galaxy SED by
adopting various physically motivated parametrisations and priors.
All these assumptions impact the recovered galaxy parameters \citep[e.g.,][]{carnall2019, leja2019, Sandles2022MNRAS.515.2951S}.

Another critical difference between the Bayesian approach and our method is
their inclusion of a nebular continuum \citep[which can be substantial
in extremely young stellar populations, e.g.][]{Byler2017,Pappalardo2021} and of differential dust attenuation
\citep{Charlot2000ApJ...539..718C}.

However, our approach also presents advantages, and has been demonstrated to
work both in the local Universe \citetext{e.g.,
\citealp{Lu2023,Zhu2023b,Zhu2023}, \citealp{barone+2018, barone+2021}}, as well
as at redshifts $z\approx1$ \citep{Cappellari2022}.
In particular, we have recently shown how \ppxf -- without the use of
any priors -- infers a realistic chemical-evolution history for local
quiescent and star-forming galaxies (Looser et~al., submitted).

In addition, and central to this work, we have shown in
Fig.~\ref{SFR_ppxf_10_100_comp} that \sfrcten is in excellent agreement with
\sfrnebten -- even though these two measurements are obtained
completely independently.
By definition, \sfrcten probes the last 10~Myr of the SFH, as
inferred through observation of the stellar continuum. 
\sfrnebten, on the other hand, is based on \Halpha, itself an indirect measure of the ionising
continuum. Because this continuum comes from massive stars with lifetimes
shorter than 10~Myr, \sfrnebten probes timescales of 3--10~Myr, comparable to \sfrcten
\citep[e.g.,][]{kennicutt1998}. The good agreement between these two observables should therefore not
be surprising. However, while \sfrcten is inferred primarily through the UV continuum redward of
\LymanAlpha, \sfrnebten measures the (highly absorbed) UV continuum \textit{blueward} of \LymanAlpha.
In principle, the correct SFR could arise `for the wrong reason', e.g., our
model neglects both the strong nebular continuum and its suppression due to
differential dust attenuation. While we cannot rule out this hypothesis, it
seems unlikely that the agreement between \sfrcten and \sfrnebten should arise
from two large systematic errors cancelling out.

In the right subplot of Fig.~\ref{SFR_ppxf_10_100_comp} we present a comparison
between $\sfrnebten$ and \sfrchund, i.e. the average SFR over 100 Myr before the epoch of
observation, as inferred from the stellar continuum. This traces the average star-formation
activity over the same timescales as empirical tracers based on rest-frame UV emission \cite[e.g.][]{Shivaei2015}. We observe again a strong correlation, but the scatter is much larger than between \sfrnebten and \sfrcten. This is already 
well known, and is commonly
interpreted as a measure of star-formation burstiness \citep[e.g.,][]{weisz+2012}.

% This is just a placeholder that I put here because too tired to write extensively. Need to read carefully Andy's message and follow some of his suggestions. If this method works, we should probably patent it.
We note that our SFR indicator may be useful to explore independently galaxy
escape fractions \fesc. Certain empirical estimators of \fesc compare a pair of
observables, the equivalent width of recombination lines (e.g., EW(\Hbeta))
and the UV slope $\beta$ to a grid of models. Our method combines the full shape
of the UV and visible continuum to infer the amount of stars of various ages.
In principle, galaxies with high \fesc should manifest as outliers in the
\sfrnebten--\sfrcten correlation, with the latter higher proportionally to \fesc.

\subsection{Stellar population age trends}

The trends of stellar age as a function of \Mstar, \DeltaMS\ and $z$ presented in Fig.~\ref{fig:Stellar-ages} provide a first measure of burstiness. A rapidly 
varying (`burstier') SFH would lead to an inconsistent
relation between \sfrnebten and mass-weighted age; for example, galaxies that 
have stopped forming stars rapidly will have low current \sfrnebten (thus
lying below the MS), but young ages (10--30~Myr), which increases the scatter 
between \DeltaMS and age itself for the population as a whole. This seems 
exactly what we observe in Fig.~\ref{fig:Stellar-ages}; while most young 
galaxies (age $\approx$ 10-30~Myr) are on and above the MS, there are several 
examples of equally young systems below the MS.

Even though the locus of the star-forming MS is likely affected by sample bias,
we see large scatter in the SFR of the youngest galaxies, larger than
the typical observational uncertainties on SFR itself.
However, for the precision of the age measurements, we may be dominated by
systematic uncertainties. A comparison with different measurements, e.g., from
Bayesian SED modelling, may help quantify what fraction of the observed scatter about the age--\DeltaMS relation is intrinsic and what is due to measurement
uncertainties.

At the same time, we still observe the clear overall trends of decreasing age 
with decreasing \Mstar, and with increasing $z$ and SFR -- which is a reassuring 
independent test of the quality of our measurements.
In particular, the trend of decreasing age with increasing $z$ is not simply
due to our truncation of the age grid to match the age of the Universe at
the redshift of each galaxy -- simply because in each redshift bin, we see
systematic age differences across the SFR--\Mstar plane. In addition, this
trend does not arise from observational bias either: if there were
significant numbers of young galaxies below the MS, they would be
systematically brighter than older galaxies at the same location on the
SFR--\Mstar plane.

The trends between age and \Mstar and SFR have been observed also in the local 
Universe \citetext{$z<0.1$; e.g.%, \citealp{blabla},
Looser et~al., submitted},
where they are interpreted as a manifestation of SFH that self correlate over
the MS timescale 1/sSFR \citep[Looser et al., submitted,][]{Tacchella2020MNRAS.497..698T}.
While this timescale is of order few Gyr in the local Universe, at high
redshift 1/sSFR is likely much shorter, of order a 100~Myr or even shorter,
so the age correlations we observe do not rule out \textit{ipso facto} 
bursty SFHs.

%are qualitatively as expected; with increasing redshift, the average age of galaxies decreases, high-mass galaxies are older than low-mass galaxies and galaxies above the MS are younger than galaxies below the MS. However, we find a large scatter about this average trend, hinting at large variability of star-formation activity over time.
%, while the \DeltaMS position of a galaxy at the epoch of observation is on average correlated to its past formation history over extended time-scales, as discussed above.
The overall picture of younger galaxies above the MS, at higher $z$ and lower
\Mstar agrees with the results we obtain from stacks of individual SFHs
in $z$--\Mstar--\DeltaMS bins (Figs.~\ref{fig:SFH_MW_bin_1}--\ref{fig:SFH_MW_bin_3}). Additionally, stacked SFH appear to show that
massive galaxies formed generally earlier and their SFR are on average
stationary or declining.
%This suggests that galaxies `jump and ride' the MS, then ride off.

Similarly to what we find for stellar age, also dust reddening trends with \Mstar, \DeltaMS\ and $z$ (Fig.~\ref{fig:Stellar-reddening}). These trends
are qualitatively as one arguably expects: quiescent galaxies and galaxies below the MS exhibit little or no evidence for dust, whereas star-bursting galaxies have significant reddening. Further, nearly all galaxies at low redshift experience some dust attenuation of the stellar continuum, while we do not observe any highly obscured objects at high-$z$ (this is unlikely to be due
to observing bias, but could be due to sample selection bias, see Sandles et~al., in~prep.). Finally, on average, massive galaxies tend to be dustier
than low-mass galaxies. These trends are derived purely from the stellar 
continuum, but they are in excellent agreement with what we infer from nebular 
recombination lines (Sandles et~al., in~prep.) -- another independent confirmation of our approach. If we assume dust attenuation roughly traces
the amount of cold gas, these trends suggest that SFR and \DeltaMS
are tightly coupled with the availability of fuel. For galaxies in the
mass range we explore, this conclusion is in excellent agreement with the
prediction of some simulations, which argue the instantaneous \DeltaMS of a
galaxy is driven by rapid gas accretion and depletion, without the `dampening'
afforded by large disc reservoirs observed in higher mass galaxies and at lower 
redshifts \citep{wang+2019}.

%for a composite view. As discussed in section \ref{ppxf_methodology} the \ppxf fits are non-parametric: \ppxf can freely choose the weighting of each individual SSP spectrum, without any priors confining the shape of the SFH. As one arguably expects, high-redshift galaxies are mostly dominated by young stellar populations. Conversely, low-redshift galaxies have older populations, with a larger fraction of stellar mass formed at early times. Within each redshift bin, galaxies below the MS have more weight in old SSPs than galaxies on the MS, while galaxies above the MS exhibit the highest mass fraction in the first two age bins. Further, we deduce a weak trend with stellar mass: massive galaxies tend to be older than dwarf galaxies.
\subsection{Bursty SFHs}

 Theoretical models agree that the scatter about the star-forming main sequence increases with
 increasing $z$, because the conditions of the primordial Universe are conducive to stronger feedback,
 which results in galaxies with more `bursty' star-formation histories \citep{Ceverino2018MNRAS.480.4842C, Lovell2022arXiv221107540L, Ma2018MNRAS.478.1694M, Faucher-Giguere2018, Tacchella2020}. However,
these models differ in how star formation and feedback are implemented, therefore they provide different quantitative predictions. A comparison with
observations is particularly useful, because burstiness is predicted both in
models with and without AGN feedback, which is thought to also affect the
evolution of low-mass galaxies \citep{Koudmani2019MNRAS.484.2047K}.
 % The physical reasons driving different levels of burstiness are
 % related to both inner mechanisms (e.g., star-formation efficiency, star-formation and AGN feedback)
 % as well as external factors (e.g., rate of gas accretion and/or mergers). Because several of these phenomena
 % require `sub-grid' physics, different studies INTRO
  % However, the level of burstiness, and the physical mechanisms responsible for this burstiness, are highly different in these models. Hence investigating bursty SFHs observationally is critical to constrain theory and to test whether the assumptions adopted by currently successful models (at low redshift) are adequate also in the infancy stages of galaxy formation. INTRO
 
Several studies have found prescriptions on how to measure
burstiness \citep[e.g., the power spectral density approach of ][]{caplar+tacchella2019} and/or how
to infer burstiness from observational data \citep{weisz+2012, faisst+2019,Wang2016, caplar+tacchella2019,speagle+2014}.

However, observational confirmation and characterisation of bursty SFHs is still challenging.
The key problem is that burstiness manifests itself as scatter -- both about
the MS, and between SFR indicators averaged over different timescales. The 
difficulty is then to extricate the information-laden scatter rooted in physical 
burstiness from the incidental contribution of measurement noise.

We argue that we overcome this difficulty in two ways. By using a non-parametric approach, we avoid biasing our solutions towards particular
SFH shapes. In practice, adding regularisation is akin to introducing a prior, 
with higher regularisation biasing the solution towards less bursty SFHs.
However, this is precisely why we use very low regularisation.
Our non-parametric approach increases the accuracy of SFHs, but comes naturally 
at the cost of decreased precision. However, JADES provides us with the means to overpower low precision measurements; its exceptionally deep spectroscopy, spans 
both the rest-frame UV and rest-frame optical, i.e., the regions of the spectrum 
dominated by stellar emission -- particularly stars with 
ages younger than 100~Myr.
We use two alternative measures of burstiness, i.e. \sfrnebten/\sfrchund
(Fig.~\ref{fig:Bursty_SFH}, left), and \sfrcten/\sfrcninety (Figs.~\ref{SFR_mass_plane_burstiness} and~\ref{fig:Bursty_SFH}, right)\footnote{We note that the inferred burstiness does not depend on the $\chi^2$-value of the fit% or the S/N of the spectrum
.}.
These are still empirical estimators which may be difficult to compare directly
to theoretical predictions.
However, compared to the classic burstiness measure \sfrnebten / \sfr{UV}{100},
we improve by using the entire information encoded in the spectrum, while \sfr{UV}{100} reduces the SFH in the last 100~Myr to a single, degenerate observable, i.e. UV luminosity.

In the future, it will be crucial to compare our results with Bayesian stellar population modelling codes,
which could potentially provide a more physically motivated reconstruction of the star formation history by incorporating complex burst patterns driven by physical expectations. In particular, Bayesian approaches could combine the posterior
probability distributions for large samples of galaxies to constrain
population-wide parameters like burstiness (e.g. using the hierarchical approach
of Wan et~al., in~prep.).

In section \ref{sec:Bursty_SFHs}, we presented observational evidence for bursty
SFHs in high-redshift and in low-mass systems, based on full spectral fitting with \ppxf (Figs.~\ref{SFR_mass_plane_burstiness} and~\ref{fig:Bursty_SFH}).
Qualitatively, this is in agreement with model expectations. As we mentioned,
at fixed stellar mass, galaxies at higher redshift had burstier histories.
Conversely, at fixed redshift, burstiness increases with decreasing stellar
mass.
The physical reason for this burstiness might be different: in high-z galaxies, burstiness is likely a result of abundant amount of cold, dense gas, and high stochasticity of the gas inflow rate combined with powerful supernovae and AGN feedback, while in low-mass galaxies, local star-bursts in short-lived giant molecular clouds might be responsible for the burstiness on short time-scales \cite{Tacchella2020}.

We observe interesting trends with redshift $z$, \DeltaMS and \Mstar: (a) we preferentially observe "bursting" systems (indicated by light colors in Fig.~\ref{SFR_mass_plane_burstiness}) at higher redshift and in low-mass systems, as one can arguably expect from observation bias \citep{Sun+2023}. (b) These systems are preferentially situated above the MS, as traced by nebular emission lines, in agreement with the interpretation that they are in a burst phase. (c) Galaxies on the MS are often in a "regular" phase (indicated by red colors). (d) Galaxies below the MS are often in "lull" phases, but also in "regular" phases, which could indicate reduced star formation activity over extended time-scales. (e) High-mass systems are often found in "regular" phases, particularly in the first two redshift bins, suggesting a more regular evolution of these systems -- again consistent with
theoretical expectations \citep[e.g.,][]{Ceverino2018MNRAS.480.4842C}.

We emphasize that the analysis presented here is a high-redshift view on SFHs. In the local Universe, the SFHs of galaxies are different. Studies of large galaxy samples at redshift $z=0$ (e.g. MaNGA \citep{Bundy2015} or SDSS \citep{Abdurro2022ApJS..259...35A}) show a strong connection between the star-formation activity of a galaxy at the epoch of observation and its past SFH. SFHs of galaxies are steadier, and their evolution is dominated by different physical processes. For example, galaxies dominantly quench through starvation \citep[][Looser et al in prep.]{Peng2015Natur.521..192P,Trussler2020,Trussler2021MNRAS.500.4469T}, with environment \citep{peng+2010} or outflows playing a secondary effect. Conversely, our analysis presented here strongly suggests, that SFHs in the young Universe are bursty.

The critical question remains whether the large \sfrcten/\sfrcninety we 
measure reflects just the tail of a much less bursty, much larger galaxy 
population. Intriguingly, the complementary approach based on photometry only
seems to reach similar conclusions \citep[][Endsley et al., submitted]{Endsley2022}. Their work uses SED modelling with \beagle of deep, nine-band NIRCam imaging from JADES, revealing evidence for "lulling" galaxy candidates.

However, to unambiguously prove the bursty SFH interpretation, galaxies in lull phases and (mini-)quenched galaxies have to be spectroscopically confirmed -- which we 
discuss next.
%As we will discuss in the next subsection, we do observe these populations in our sample, in regimes where our sample is not as strongly affected by selection bias, i.e. at slightly lower redshift and slightly higher masses. 

\subsection{Galaxies in "lulls" and (mini-)quenching at high redshift}

 In section \ref{Lull_normal_examples}, we show that the extraordinary
 depth and data quality of JADES are capable of probing well below the
 starburst regime. Indeed, we have shown examples of both galaxies on
 the MS (i.e., "regular" phase, Fig.~\ref{normal_example}) as well as below the MS
 (in "lull" phases; Fig.~\ref{lull_example}).
 We further show that JADES is capable of probing galaxies that are
 formally below the redshift-dependent quiescent threshold
 in all three redshift bins (Fig.~\ref{fig:Stellar-ages}). These galaxies show no
 evidence for emission lines \citetext{Fig.~\ref{fig:Mini-quenched} and
 \citealp{Looser+2023}}, and are experiencing (potentially short-lived)
 phases of quiescence.
 The fact that these galaxies are present in our sample argues against
 our findings about burstiness being a result of limited sensitivity.
 However, the fraction of lull or (mini-)quenched galaxies is clearly
 lower than that of star-burst systems (Fig.~\ref{fig:Stellar-ages}).
 This could ostensibly be the undesired outcome of the initial sample selection, which prioritised objects with
 high-confidence photometric redshifts. In practice, these amount to
 galaxies that have stronger broad-band drops, due in turn to either
Lyman continuum drop or to high-equivalent-width nebular emission -- both
of which generally associated with high SFR \citetext{Bunker et~al.,
submitted}. The difficulty of an unbiased sample selection is highlighted
by the complementary approach based on JADES photometry (Endsley et~al., submitted), which finds very similar results (i.e,. an overabundance of
starburst galaxies) based on a NIRCam-selected sample.
Indeed, back to our spectroscopic sample, the fact that the only mini-quenched 
galaxy at $z>5$ is relatively massive and young 
\citetext{$\Mstar = 5 \times 10^8$~\MSun, $10^8$~yr;
\citealp{Looser+2023}} suggests that -- in this redshift range, we are
limited by the depth of JADES. This is again supported by the fact that,
in the intermediate-redshift bin, we are able to confirm a mini-quenched galaxy 
with $\Mstar = 8 \times 10^7$~\MSun and age $10^{7.9}$~yr -- clearly in the
low-mass regime where strong feedback is expected to trigger short-lived
mini-quenching \citep{Ceverino2018MNRAS.480.4842C,Ma2018MNRAS.478.1694M,Dome2023}.

According to models, the burstiness of star formation should only increase
probing masses $\Mstar \approx 10^6$~\MSun. The galaxies we present here, 
together with those presented in recent works \citep{Strait2023, Looser+2023}, 
show that we are finally beginning to probe the obverse face of the
burstiness phenomenon. However,
a more quantitative understanding will probably require the expensive 
combination of larger samples and/or even deeper observations. In particular, 
separating the degenerate effects of mass and redshift will only be feasible 
with thousands of objects probing effectively the parameter space.

 Nonetheless, the analysis presented in this work is strong evidence that galaxies at high-redshift are bursty, and endure these "lulling" and (mini-)quenched phases. In the next section, we discuss the effects of observation bias in more detail.

\section{Summary and conclusions} \label{sec:Summary}

In this work, we combine the non-parametric approach of the \ppxf software with the marvellous depth of \jwst/NIRSpec MSA spectroscopy to offer a view on the
star-formation history (SFH) of low- to intermediate-mass galaxies ($10^{6}<\Mstar<10^{9.5}$~\MSun) at cosmic dawn, between redshifts $0.6\lesssim z \lesssim 11$.

The key results of this paper are as follows:
\begin{itemize}
    \item The correlation of (mass-weighted stellar-population)
    age with \Mstar, $z$ and SFR existed already well before the local Universe
    and even earlier than Cosmic Noon, at redshifts as high as $5<z<11$
    (Fig.~\ref{fig:Stellar-ages}). All else being equal, age increases with increasing \Mstar and decreases with increasing $z$ and with increasing distance from the
    star-forming main sequence, \DeltaMS.
    We find consistent trends in the stellar populations in stacks of SFHs in $z$--$\Mstar$--\DeltaMS bins.
    The existence of these correlations is unlikely to result
    from sample or observation bias, and argues for the SFH of galaxies being
    correlated on timescales comparable to the main-sequence timescale
    of 1/sSFR (where sSFR is the specific star-formation rate).
    \item However, the trends between age, M$_*$ and SFR, have large scatter, with examples of young stellar populations also below the Main Sequence, which gives a first indication of bursty star formation.
    \item We introduce and validate a short-timescale, continuum based SFR
    indicator, averaged over 10~Myr (\sfrcten). We compare \sfrcten to
    the average over the last 10--100 Myr (\sfrcninety) as an estimate of
    SFH burstiness.
    \item By using these parameters, we present additional observational evidence that the SFHs of high-redshift and low-mass galaxies are bursty. Specifically, we use \sfrcten/\sfrcninety to investigate burstiness in the SFR-mass plane and as a function of redshift, and find that  high redshift and low-mass galaxies have particularly bursty SFHs,
while more massive and lower-redshift systems evolve more steadily. .
    \item We report the discovery of another low-mass galaxy in the fourth phase, which we call: "(mini-)quenched", at redshift $z=4.4$. This
    galaxy lies well within the mass regime where numerical simulations predict star-formation being dominated by short and intense bursts. Therefore, the
    quiescence of this galaxy might be only transient, as discussed in \cite{Looser+2023}.
    \item We argue that we see most targets at the observability frontier, i.e. at the highest redshifts and the lowest mass systems, preferentially in "bursting" phases. Their more regular, lulling and (mini-)quenched counterparts are likely at the bottom edge of the observability window at the epoch of observation, even for \jwst.
     \item Finally, we use the stellar $E(B-V)$ as a proxy for the amount of dust and find that 
     $E(B-V)$ increases with
    increasing \Mstar and \DeltaMS, possibly as a result of the correlation
    between dust and gas mass, and gas mass and SFR in step.
    We find that $E(B-V)$ decreases at the highest redshifts, although
     most galaxies at $z>5$ still have some dust. 
\end{itemize}

However, this is only the beginning of the investigation of stellar populations and bursty SFHs and (mini-)quenching at high redshift with galaxy population samples observed with \jwst: larger statistical samples of high-S/N galaxy spectra will enable the investigation and quantification of selection effects, which are key to this kind of analysis, and the quantification of various physical aspects of stellar populations and bursty SFHs, like duty cycles, oscillation times, short- and long-term variability, etc., e.g. in the framework of the power spectral density \citep[PSD][]{Tacchella2020}. And a detailed comparison to numerical cosmological simulations will be crucial to analyse the complex interplay of physical mechanisms contributing to making SFHs bursty. Upcoming observations with \jwst will provide such a sample, and will continue to reveal the physics processes which shape the observed differing assembly histories of galaxies in the early Universe.

\begin{acknowledgements}
TJL, FDE, RM, LS, WB, LS, JS and JW acknowledge support by the Science and Technology Facilities Council (STFC) and by the ERC through Advanced Grant 695671 “QUENCH”. TJL and ALD acknowledge support by the STFC Center for Doctoral Training in data intensive science program. RM also acknowledges funding from a research professorship from the Royal Society. ECL acknowledges support of an STFC Webb Fellowship (ST/W001438/1). SA and BRP acknowledge support from Grant PID2021-127718NB-I00 funded by the Spanish Ministry of Science and Innovation/State Agency of Research (MICIN/AEI/ 10.13039/501100011033). SC acknowledges support by European Union’s HE ERC Starting Grant No. 101040227 - WINGS. 
AJB, and JC acknowledge funding from the "FirstGalaxies" Advanced Grant from the European Research Council (ERC) under the European Union’s Horizon 2020 research and innovation program (Grant agreement No. 789056). 
H{\"U} gratefully acknowledges support by the Isaac Newton Trust and by the Kavli Foundation through a Newton-Kavli Junior Fellowship. 
JW further acknowledges support from the Fondation MERAC.
KB acknowledges support by the Australian Research Council Centre of Excellence for All Sky Astrophysics in 3 Dimensions (ASTRO 3D), through project number CE170100013. The Cosmic Dawn Center (DAWN) is funded by the Danish National Research Foundation under grant no.140. DJE is supported as a Simons Investigator.
ALD thanks the Cambridge Harding postgraduate program.
DJE BDJ and BER acknowledge support by the JWST/NIRCam contract to the University of Arizona, NAS5-02015. 
RS acknowledges support from a STFC Ernest Rutherford Fellowship (ST/S004831/1). The research of CCW is supported by NOIRLab, which is managed by the Association of Universities for Research in Astronomy (AURA) under a cooperative agreement with the National Science Foundation.
\end{acknowledgements}

% WARNING
%-------------------------------------------------------------------
% Please note that we have included the references to the file aa.dem in
% order to compile it, but we ask you to:
%
% - use BibTeX with the regular commands:
%   \bibliographystyle{aa} % style aa.bst
%   \bibliography{Yourfile} % your references Yourfile.bib
%
% - join the .bib files when you upload your source files
%-------------------------------------------------------------------

\bibliographystyle{aa}
\bibliography{Bib_TL}

\begin{thebibliography}{84}
\expandafter\ifx\csname natexlab\endcsname\relax\def\natexlab#1{#1}\fi

\bibitem[{{Abdurro'uf} {et~al.}(2022){Abdurro'uf}, {Accetta}, {Aerts}, {Silva
  Aguirre}, {Ahumada}, {Ajgaonkar}, {Filiz Ak}, {Alam}, {Allende Prieto},
  {Almeida}, {Anders}, {Anderson}, {Andrews}, {Anguiano}, {Aquino-Ort{\'\i}z},
  {Arag{\'o}n-Salamanca}, {Argudo-Fern{\'a}ndez}, {Ata}, {Aubert},
  {Avila-Reese}, {Badenes}, {Barb{\'a}}, {Barger}, {Barrera-Ballesteros},
  {Beaton}, {Beers}, {Belfiore}, {Bender}, {Bernardi}, {Bershady}, {Beutler},
  {Bidin}, {Bird}, {Bizyaev}, {Blanc}, {Blanton}, {Boardman}, {Bolton},
  {Boquien}, {Borissova}, {Bovy}, {Brandt}, {Brown}, {Brownstein}, {Brusa},
  {Buchner}, {Bundy}, {Burchett}, {Bureau}, {Burgasser}, {Cabang}, {Campbell},
  {Cappellari}, {Carlberg}, {Wanderley}, {Carrera}, {Cash}, {Chen}, {Chen},
  {Cherinka}, {Chiappini}, {Choi}, {Chojnowski}, {Chung}, {Clerc}, {Cohen},
  {Comerford}, {Comparat}, {da Costa}, {Covey}, {Crane}, {Cruz-Gonzalez},
  {Culhane}, {Cunha}, {Dai}, {Damke}, {Darling}, {Davidson}, {Davies},
  {Dawson}, {De Lee}, {Diamond-Stanic}, {Cano-D{\'\i}az}, {S{\'a}nchez},
  {Donor}, {Duckworth}, {Dwelly}, {Eisenstein}, {Elsworth}, {Emsellem},
  {Eracleous}, {Escoffier}, {Fan}, {Farr}, {Feng}, {Fern{\'a}ndez-Trincado},
  {Feuillet}, {Filipp}, {Fillingham}, {Frinchaboy}, {Fromenteau}, {Galbany},
  {Garc{\'\i}a}, {Garc{\'\i}a-Hern{\'a}ndez}, {Ge}, {Geisler}, {Gelfand},
  {G{\'e}ron}, {Gibson}, {Goddy}, {Godoy-Rivera}, {Grabowski}, {Green},
  {Greener}, {Grier}, {Griffith}, {Guo}, {Guy}, {Hadjara}, {Harding},
  {Hasselquist}, {Hayes}, {Hearty}, {Hern{\'a}ndez}, {Hill}, {Hogg},
  {Holtzman}, {Horta}, {Hsieh}, {Hsu}, {Hsu}, {Huber}, {Huertas-Company},
  {Hutchinson}, {Hwang}, {Ibarra-Medel}, {Chitham}, {Ilha}, {Imig}, {Jaekle},
  {Jayasinghe}, {Ji}, {Johnson}, {Jones}, {J{\"o}nsson}, {Katkov}, {Khalatyan},
  {Kinemuchi}, {Kisku}, {Knapen}, {Kneib}, {Kollmeier}, {Kong}, {Kounkel},
  {Kreckel}, {Krishnarao}, {Lacerna}, {Lane}, {Langgin}, {Lavender}, {Law},
  {Lazarz}, {Leung}, {Leung}, {Lewis}, {Li}, {Li}, {Lian}, {Liang}, {Lin},
  {Lin}, {Lin}, {Lintott}, {Long}, {Longa-Pe{\~n}a}, {L{\'o}pez-Cob{\'a}},
  {Lu}, {Lundgren}, {Luo}, {Mackereth}, {de la Macorra}, {Mahadevan},
  {Majewski}, {Manchado}, {Mandeville}, {Maraston}, {Margalef-Bentabol},
  {Masseron}, {Masters}, {Mathur}, {McDermid}, {Mckay}, {Merloni},
  {Merrifield}, {Meszaros}, {Miglio}, {Di Mille}, {Minniti}, {Minsley},
  {Monachesi}, {Moon}, {Mosser}, {Mulchaey}, {Muna}, {Mu{\~n}oz}, {Myers},
  {Myers}, {Nadathur}, {Nair}, {Nandra}, {Neumann}, {Newman}, {Nidever},
  {Nikakhtar}, {Nitschelm}, {O'Connell}, {Garma-Oehmichen}, {Luan Souza de
  Oliveira}, {Olney}, {Oravetz}, {Ortigoza-Urdaneta}, {Osorio}, {Otter},
  {Pace}, {Padilla}, {Pan}, {Pan}, {Parikh}, {Parker}, {Peirani}, {Pe{\~n}a
  Ram{\'\i}rez}, {Penny}, {Percival}, {Perez-Fournon}, {Pinsonneault},
  {Poidevin}, {Poovelil}, {Price-Whelan}, {B{\'a}rbara de Andrade Queiroz},
  {Raddick}, {Ray}, {Rembold}, {Riddle}, {Riffel}, {Riffel}, {Rix}, {Robin},
  {Rodr{\'\i}guez-Puebla}, {Roman-Lopes}, {Rom{\'a}n-Z{\'u}{\~n}iga}, {Rose},
  {Ross}, {Rossi}, {Rubin}, {Salvato}, {S{\'a}nchez}, {S{\'a}nchez-Gallego},
  {Sanderson}, {Santana Rojas}, {Sarceno}, {Sarmiento}, {Sayres}, {Sazonova},
  {Schaefer}, {Schiavon}, {Schlegel}, {Schneider}, {Schultheis}, {Schwope},
  {Serenelli}, {Serna}, {Shao}, {Shapiro}, {Sharma}, {Shen}, {Shetrone}, {Shu},
  {Simon}, {Skrutskie}, {Smethurst}, {Smith}, {Sobeck}, {Spoo}, {Sprague},
  {Stark}, {Stassun}, {Steinmetz}, {Stello}, {Stone-Martinez},
  {Storchi-Bergmann}, {Stringfellow}, {Stutz}, {Su}, {Taghizadeh-Popp},
  {Talbot}, {Tayar}, {Telles}, {Teske}, {Thakar}, {Theissen}, {Tkachenko},
  {Thomas}, {Tojeiro}, {Hernandez Toledo}, {Troup}, {Trump}, {Trussler},
  {Turner}, {Tuttle}, {Unda-Sanzana}, {V{\'a}zquez-Mata}, {Valentini},
  {Valenzuela}, {Vargas-Gonz{\'a}lez}, {Vargas-Maga{\~n}a}, {Alfaro},
  {Villanova}, {Vincenzo}, {Wake}, {Warfield}, {Washington}, {Weaver},
  {Weijmans}, {Weinberg}, {Weiss}, {Westfall}, {Wild}, {Wilde}, {Wilson},
  {Wilson}, {Wilson}, {Wolf}, {Wood-Vasey}, {Yan}, {Zamora}, {Zasowski},
  {Zhang}, {Zhao}, {Zheng}, {Zheng}, \& {Zhu}}]{Abdurro2022ApJS..259...35A}
{Abdurro'uf}, {Accetta}, K., {Aerts}, C., {et~al.} 2022, \apjs, 259, 35

\bibitem[{{Barone} {et~al.}(2018){Barone}, {D'Eugenio}, {Colless}, {Scott},
  {van de Sande}, {Bland-Hawthorn}, {Brough}, {Bryant}, {Cortese}, {Croom},
  {Foster}, {Goodwin}, {Konstantopoulos}, {Lawrence}, {Lorente}, {Medling},
  {Owers}, \& {Richards}}]{barone+2018}
{Barone}, T.~M., {D'Eugenio}, F., {Colless}, M., {et~al.} 2018, \apj, 856, 64

\bibitem[{{Barone} {et~al.}(2021){Barone}, {D'Eugenio}, {Scott}, {Colless},
  {Vaughan}, {van der Wel}, {Fraser-McKelvie}, {de Graaff}, {van de Sande},
  {Wu}, {Bezanson}, {Brough}, {Bell}, {Croom}, {Cortese}, {Driver}, {Gallazzi},
  {Muzzin}, {Sobral}, {Bland-Hawthorn}, {Bryant}, {Goodwin}, {Lawrence},
  {Lorente}, \& {Owers}}]{barone+2021}
{Barone}, T.~M., {D'Eugenio}, F., {Scott}, N., {et~al.} 2021, arXiv e-prints,
  arXiv:2107.01054

\bibitem[{Bruzual \& Charlot(2003)}]{bruzual_stellar_2003}
Bruzual, G. \& Charlot, S. 2003, \mnras, 344, 1000

\bibitem[{{Bundy} {et~al.}(2015){Bundy}, {Bershady}, {Law}, {Yan}, {Drory},
  {MacDonald}, {Wake}, {Cherinka}, {S{\'a}nchez-Gallego}, {Weijmans}, {Thomas},
  {Tremonti}, {Masters}, {Coccato}, {Diamond-Stanic}, {Arag{\'o}n-Salamanca},
  {Avila-Reese}, {Badenes}, {Falc{\'o}n-Barroso}, {Belfiore}, {Bizyaev},
  {Blanc}, {Bland-Hawthorn}, {Blanton}, {Brownstein}, {Byler}, {Cappellari},
  {Conroy}, {Dutton}, {Emsellem}, {Etherington}, {Frinchaboy}, {Fu}, {Gunn},
  {Harding}, {Johnston}, {Kauffmann}, {Kinemuchi}, {Klaene}, {Knapen},
  {Leauthaud}, {Li}, {Lin}, {Maiolino}, {Malanushenko}, {Malanushenko}, {Mao},
  {Maraston}, {McDermid}, {Merrifield}, {Nichol}, {Oravetz}, {Pan}, {Parejko},
  {Sanchez}, {Schlegel}, {Simmons}, {Steele}, {Steinmetz}, {Thanjavur},
  {Thompson}, {Tinker}, {van den Bosch}, {Westfall}, {Wilkinson}, {Wright},
  {Xiao}, \& {Zhang}}]{Bundy2015}
{Bundy}, K., {Bershady}, M.~A., {Law}, D.~R., {et~al.} 2015, APJ, 798, 7

\bibitem[{{Byler} {et~al.}(2017){Byler}, {Dalcanton}, {Conroy}, \&
  {Johnson}}]{Byler2017}
{Byler}, N., {Dalcanton}, J.~J., {Conroy}, C., \& {Johnson}, B.~D. 2017, \apj,
  840, 44

\bibitem[{{Calzetti} {et~al.}(2000){Calzetti}, {Armus}, {Bohlin}, {Kinney},
  {Koornneef}, \& {Storchi-Bergmann}}]{Calzetti2000}
{Calzetti}, D., {Armus}, L., {Bohlin}, R.~C., {et~al.} 2000, APJ, 533, 682

\bibitem[{{Calzetti} {et~al.}(1994){Calzetti}, {Kinney}, \&
  {Storchi-Bergmann}}]{Calzetti1994}
{Calzetti}, D., {Kinney}, A.~L., \& {Storchi-Bergmann}, T. 1994, \apj, 429, 582

\bibitem[{{Caplar} \& {Tacchella}(2019)}]{caplar+tacchella2019}
{Caplar}, N. \& {Tacchella}, S. 2019, \mnras, 487, 3845

\bibitem[{{Cappellari}(2017)}]{Cappellari2017}
{Cappellari}, M. 2017, \mnras, 466, 798

\bibitem[{{Cappellari}(2022)}]{Cappellari2022}
{Cappellari}, M. 2022, arXiv e-prints, arXiv:2208.14974

\bibitem[{{Caputi} {et~al.}(2007){Caputi}, {Lagache}, {Yan}, {Dole},
  {Bavouzet}, {Le Floc'h}, {Choi}, {Helou}, \& {Reddy}}]{Caputi2007}
{Caputi}, K.~I., {Lagache}, G., {Yan}, L., {et~al.} 2007, \apj, 660, 97

\bibitem[{{Carnall} {et~al.}(2019{\natexlab{a}}){Carnall}, {Leja}, {Johnson},
  {McLure}, {Dunlop}, \& {Conroy}}]{carnall2019}
{Carnall}, A.~C., {Leja}, J., {Johnson}, B.~D., {et~al.} 2019{\natexlab{a}},
  \apj, 873, 44

\bibitem[{{Carnall} {et~al.}(2023){Carnall}, {McLeod}, {McLure}, {Dunlop},
  {Begley}, {Cullen}, {Donnan}, {Hamadouche}, {Jewell}, {Jones}, {Pollock}, \&
  {Wild}}]{carnall+2023b}
{Carnall}, A.~C., {McLeod}, D.~J., {McLure}, R.~J., {et~al.} 2023, \mnras, 520,
  3974

\bibitem[{{Carnall} {et~al.}(2019{\natexlab{b}}){Carnall}, {McLure}, {Dunlop},
  {Cullen}, {McLeod}, {Wild}, {Johnson}, {Appleby}, {Dav{\'e}}, {Amorin},
  {Bolzonella}, {Castellano}, {Cimatti}, {Cucciati}, {Gargiulo}, {Garilli},
  {Marchi}, {Pentericci}, {Pozzetti}, {Schreiber}, {Talia}, \&
  {Zamorani}}]{Carnall2019MNRAS.490..417C}
{Carnall}, A.~C., {McLure}, R.~J., {Dunlop}, J.~S., {et~al.}
  2019{\natexlab{b}}, \mnras, 490, 417

\bibitem[{{Carnall} {et~al.}(2018){Carnall}, {McLure}, {Dunlop}, \&
  {Dav{\'e}}}]{Carnall2018}
{Carnall}, A.~C., {McLure}, R.~J., {Dunlop}, J.~S., \& {Dav{\'e}}, R. 2018,
  \mnras, 480, 4379

\bibitem[{{Ceverino} {et~al.}(2021){Ceverino}, {Hirschmann}, {Klessen},
  {Glover}, {Charlot}, \& {Feltre}}]{Ceverino2021}
{Ceverino}, D., {Hirschmann}, M., {Klessen}, R.~S., {et~al.} 2021, \mnras, 504,
  4472

\bibitem[{{Ceverino} {et~al.}(2018){Ceverino}, {Klessen}, \&
  {Glover}}]{Ceverino2018MNRAS.480.4842C}
{Ceverino}, D., {Klessen}, R.~S., \& {Glover}, S. C.~O. 2018, \mnras, 480, 4842

\bibitem[{Chabrier(2003)}]{chabrier_galactic_2003}
Chabrier, G. 2003, \pasp, 115, 763

\bibitem[{{Charlot} \& {Fall}(2000)}]{Charlot2000ApJ...539..718C}
{Charlot}, S. \& {Fall}, S.~M. 2000, \apj, 539, 718

\bibitem[{{Chevallard} \& {Charlot}(2016)}]{chevallard_beagle_2016}
{Chevallard}, J. \& {Charlot}, S. 2016, \mnras, 462, 1415

\bibitem[{{Choi} {et~al.}(2016){Choi}, {Dotter}, {Conroy}, {Cantiello},
  {Paxton}, \& {Johnson}}]{Choi2016}
{Choi}, J., {Dotter}, A., {Conroy}, C., {et~al.} 2016, \apj, 823, 102

\bibitem[{{Conroy} {et~al.}(2019){Conroy}, {Naidu}, {Zaritsky}, {Bonaca},
  {Cargile}, {Johnson}, \& {Caldwell}}]{Conroy2019}
{Conroy}, C., {Naidu}, R.~P., {Zaritsky}, D., {et~al.} 2019, \apj, 887, 237

\bibitem[{{Curti} {et~al.}(2023){Curti}, {Maiolino}, {Carniani}, {D'Eugenio},
  {Chevallard}, {Curtis-Lake}, {Looser}, {Scholtz}, {{\"U}bler}, {Witstok},
  {Cameron}, {Charlot}, {Laseter}, {Sandles}, {Arribas}, {Bunker}, {Giardino},
  {Maseda}, {Rawle}, {Rodr{\'\i}guez Del Pino}, {Smit}, {Willott},
  {Eisenstein}, {Hausen}, {Johnson}, {Rieke}, {Robertson}, {Tacchella},
  {Williams}, {Willmer}, {Baker}, {Bhatawdekar}, {Boyett}, {Egami}, {Helton},
  {Ji}, {Kumari}, {Shivaei}, \& {Sun}}]{Curti2023}
{Curti}, M., {Maiolino}, R., {Carniani}, S., {et~al.} 2023, arXiv e-prints,
  arXiv:2304.08516

\bibitem[{{D{\'\i}az-Santos} {et~al.}(2017){D{\'\i}az-Santos}, {Armus},
  {Charmandaris}, {Lu}, {Stierwalt}, {Stacey}, {Malhotra}, {van der Werf},
  {Howell}, {Privon}, {Mazzarella}, {Goldsmith}, {Murphy}, {Barcos-Mu{\~n}oz},
  {Linden}, {Inami}, {Larson}, {Evans}, {Appleton}, {Iwasawa}, {Lord},
  {Sanders}, \& {Surace}}]{Diaz_Santos2017}
{D{\'\i}az-Santos}, T., {Armus}, L., {Charmandaris}, V., {et~al.} 2017, \apj,
  846, 32

\bibitem[{{Dome} {et~al.}(2023){Dome}, {Tacchella}, {Fialkov}, {Dekel},
  {Ginzburg}, {Lapiner}, \& {Looser}}]{Dome2023}
{Dome}, T., {Tacchella}, S., {Fialkov}, A., {et~al.} 2023, arXiv e-prints,
  arXiv:2305.07066

\bibitem[{{Endsley} {et~al.}(2021){Endsley}, {Stark}, {Chevallard}, \&
  {Charlot}}]{Endsley+2021}
{Endsley}, R., {Stark}, D.~P., {Chevallard}, J., \& {Charlot}, S. 2021, \mnras,
  500, 5229

\bibitem[{{Endsley} {et~al.}(2022){Endsley}, {Stark}, {Whitler}, {Topping},
  {Chen}, {Plat}, {Chisholm}, \& {Charlot}}]{Endsley2022}
{Endsley}, R., {Stark}, D.~P., {Whitler}, L., {et~al.} 2022, arXiv e-prints,
  arXiv:2208.14999

\bibitem[{{Faisst} {et~al.}(2019){Faisst}, {Capak}, {Emami}, {Tacchella}, \&
  {Larson}}]{faisst+2019}
{Faisst}, A.~L., {Capak}, P.~L., {Emami}, N., {Tacchella}, S., \& {Larson},
  K.~L. 2019, \apj, 884, 133

\bibitem[{{Faucher-Gigu{\`e}re}(2018)}]{Faucher-Giguere2018}
{Faucher-Gigu{\`e}re}, C.-A. 2018, \mnras, 473, 3717

\bibitem[{{Ferruit} {et~al.}(2022){Ferruit}, {Jakobsen}, {Giardino}, {Rawle},
  {Alves de Oliveira}, {Arribas}, {Beck}, {Birkmann}, {B{\"o}ker}, {Bunker},
  {Charlot}, {de Marchi}, {Franx}, {Henry}, {Karakla}, {Kassin}, {Kumari},
  {L{\'o}pez-Caniego}, {L{\"u}tzgendorf}, {Maiolino}, {Manjavacas}, {Marston},
  {Moseley}, {Muzerolle}, {Pirzkal}, {Rauscher}, {Rix}, {Sabbi}, {Sirianni},
  {te Plate}, {Valenti}, {Willott}, \& {Zeidler}}]{Ferruit2022}
{Ferruit}, P., {Jakobsen}, P., {Giardino}, G., {et~al.} 2022, \aap, 661, A81

\bibitem[{{Gallazzi} {et~al.}(2014){Gallazzi}, {Bell}, {Zibetti}, {Brinchmann},
  \& {Kelson}}]{Gallazzi2014}
{Gallazzi}, A., {Bell}, E.~F., {Zibetti}, S., {Brinchmann}, J., \& {Kelson},
  D.~D. 2014, \apj, 788, 72

\bibitem[{{Gardner} {et~al.}(2023){Gardner}, {Mather}, {Abbott}, {Abell},
  {Abernathy}, {Abney}, {Abraham}, {Abraham}, {Abul-Huda}, {Acton}, {Adams},
  {Adams}, {Adler}, {Adriaensen}, {Aguilar}, {Ahmed}, {Ahmed}, {Ahmed},
  {Albat}, {Albert}, {Alberts}, {Aldridge}, {Marsha Allen}, {Allen},
  {Altenburg}, {Altunc}, {Alvarez}, {{\'A}lvarez-M{\'a}rquez}, {Alves de
  Oliveira}, {Ambrose}, {Anandakrishnan}, {Andersen}, {Anderson}, {Anderson},
  {Anderson}, {Anderson}, {Aprea}, {Archer}, {Arenberg}, {Argyriou}, {Arribas},
  {Artigau}, {Arvai}, {Atcheson}, {Atkinson}, {Averbukh}, {Aymergen},
  {Bacinski}, {Baggett}, {Bagnasco}, {Baker}, {Balzano}, {Banks}, {Baran},
  {Barker}, {Barrett}, {Barringer}, {Barto}, {Bast}, {Baudoz}, {Baum},
  {Beatty}, {Beaulieu}, {Bechtold}, {Beck}, {Beddard}, {Beichman}, {Bellagama},
  {Bely}, {Berger}, {Bergeron}, {Darveau-Bernier}, {Bertch}, {Beskow}, {Betz},
  {Biagetti}, {Birkmann}, {Bjorklund}, {Blackwood}, {Blazek}, {Blossfeld},
  {Bluth}, {Boccaletti}, {Boegner}, {Bohlin}, {Boia}, {B{\"o}ker},
  {Bonaventura}, {Bond}, {Bosley}, {Boucarut}, {Bouchet}, {Bouwman}, {Bower},
  {Bowers}, {Bowers}, {Boyce}, {Boyer}, {Boyer}, {Boyer}, {Boyer}, {Bradley},
  {Brady}, {Brandl}, {Brannen}, {Breda}, {Bremmer}, {Brennan}, {Bresnahan},
  {Bright}, {Broiles}, {Bromenschenkel}, {Brooks}, {Brooks}, {Brown}, {Brown},
  {Brown}, {Bruce}, {Bryson}, {Bujanda}, {Bullock}, {Bunker}, {Bureo}, {Burt},
  {Bush}, {Bushouse}, {Bussman}, {Cabaud}, {Cale}, {Calhoon}, {Calvani},
  {Canipe}, {Caputo}, {Cara}, {Carey}, {Case}, {Cesari}, {Cetorelli}, {Chance},
  {Chandler}, {Chaney}, {Chapman}, {Charlot}, {Chayer}, {Cheezum}, {Chen},
  {Chen}, {Cherinka}, {Chichester}, {Chilton}, {Chittiraibalan}, {Clampin},
  {Clark}, {Clark}, {Clark}, {Claybrooks}, {Cleveland}, {Cohen}, {Cohen},
  {Col{\'o}n}, {Coleman}, {Colina}, {Comber}, {Comeau}, {Comer}, {Conde Reis},
  {Connolly}, {Conroy}, {Contos}, {Contreras}, {Cook}, {Cooper}, {Aviva
  Cooper}, {Correia}, {Correnti}, {Cossou}, {Costanza}, {Coulais}, {Cox},
  {Coyle}, {Cracraft}, {Noriega-Crespo}, {Crew}, {Curtis}, {Cusveller}, {Da
  Costa Maciel}, {Dailey}, {Daugeron}, {Davidson}, {Davies}, {Davis}, {Davis},
  {Day}, {de Chambure}, {de Jong}, {De Marchi}, {Dean}, {Decker}, {Delisa},
  {Dell}, {Dellagatta}, {Dembinska}, {Demosthenes}, {Dencheva}, {Deneu},
  {DePriest}, {Deschenes}, {Dethienne}, {Hunor Detre}, {Izela Diaz}, {Dicken},
  {DiFelice}, {Dillman}, {Disharoon}, {van Dishoeck}, {Dixon}, {Doggett},
  {Dominguez}, {Donaldson}, {Doria-Warner}, {Dos Santos}, {Doty}, {Douglas},
  {Doyon}, {Dressler}, {Driggers}, {Driggers}, {Dunn}, {DuPrie}, {Dupuis},
  {Durning}, {Dutta}, {Earl}, {Eccleston}, {Ecobichon}, {Egami},
  {Ehrenwinkler}, {Eisenhamer}, {Eisenhower}, {Eisenstein}, {El Hamel}, {Elie},
  {Elliott}, {Elliott}, {Engesser}, {Espinoza}, {Etienne}, {Etxaluze}, {Evans},
  {Fabreguettes}, {Falcolini}, {Falini}, {Fatig}, {Feeney}, {Feinberg}, {Fels},
  {Ferdous}, {Ferguson}, {Ferrarese}, {Ferreira}, {Ferruit}, {Ferry},
  {Filippazzo}, {Firre}, {Fix}, {Flagey}, {Flanagan}, {Fleming}, {Florian},
  {Flynn}, {Foiadelli}, {Fontaine}, {Fontanella}, {Forshay}, {Fortner}, {Fox},
  {Framarini}, {Francisco}, {Franck}, {Franx}, {Franz}, {Friedman}, {Friend},
  {Frost}, {Fu}, {Fullerton}, {Gaillard}, {Galkin}, {Gallagher}, {Galyer},
  {Garc{\'\i}a Mar{\'\i}n}, {Gardner}, {Garland}, {Garrett}, {Gasman},
  {G{\'a}sp{\'a}r}, {Gastaud}, {Gaudreau}, {Gauthier}, {Geers}, {Geithner},
  {Gennaro}, {Gerber}, {Gereau}, {Giampaoli}, {Giardino}, {Gibbons}, {Gilbert},
  {Gilman}, {Girard}, {Giuliano}, {Gkountis}, {Glasse}, {Glassmire}, {Glauser},
  {Glazer}, {Goldberg}, {Golimowski}, {Gonzaga}, {Gordon}, {Gordon},
  {Goudfrooij}, {Gough}, {Graham}, {Grau}, {Green}, {Greene}, {Greene},
  {Greenfield}, {Greenhouse}, {Greve}, {Greville}, {Grimaldi}, {Groe},
  {Groebner}, {Grumm}, {Grundy}, {G{\"u}del}, {Guillard}, {Guldalian}, {Gunn},
  {Gurule}, {Meyer Gutman}, {Guy}, {Guyot}, {Hack}, {Haderlein}, {Hagan},
  {Hagedorn}, {Hainline}, {Haley}, {Hami}, {Hamilton}, {Hammann}, {Hammel},
  {Hanley}, {Hansen}, {Hardy}, {Harnisch}, {Harr}, {Harris}, {Hart}, {Hartig},
  {Hasan}, {Hashim}, {Hashimoto}, {Haskins}, {Hawkins}, {Hayden}, {Hayden},
  {Healy}, {Hecht}, {Heeg}, {Hejal}, {Helm}, {Hengemihle}, {Henning}, {Henry},
  {Henry}, {Henshaw}, {Hernandez}, {Herrington}, {Heske}, {Hesman}, {Hickey},
  {Hilbert}, {Hines}, {Hinz}, {Hirsch}, {Hitcho}, {Hodapp}, {Hodge}, {Hoffman},
  {Holfeltz}, {Holler}, {Hoppa}, {Horner}, {Howard}, {Howard}, {Huber},
  {Hunkeler}, {Hunter}, {Hunter}, {Hurd}, {Hurst}, {Hutchings}, {Hylan},
  {Ilinca Ignat}, {Illingworth}, {Irish}, {Isaacs}, {Jackson}, {Jaffe},
  {Jahic}, {Jahromi}, {Jakobsen}, {James}, {James}, {James}, {Jamieson},
  {Jandra}, {Jayawardhana}, {Jedrzejewski}, {Jeffers}, {Jensen}, {Joanne},
  {Johns}, {Johnson}, {Johnson}, {Johnson}, {Johnson}, {Johnson}, {Johnson},
  {Johnstone}, {Jollet}, {Jones}, {Jones}, {Jones}, {Jones}, {Jones}, {Jordan},
  {Jordan}, {Jue}, {Jurkowski}, {Justis}, {Justtanont}, {Kaleida}, {Kalirai},
  {Cabrales Kalmanson}, {Kaltenegger}, {Kammerer}, {Kan}, {Childs Kanarek},
  {Kao}, {Karakla}, {Karl}, {Kassin}, {Kauffman}, {Kavanagh}, {Kelley},
  {Kelly}, {Kendrew}, {Kennedy}, {Kenny}, {Keski-Kuha}, {Keyes}, {Khan},
  {Kidwell}, {Kimble}, {King}, {King}, {Kinzel}, {Kirk}, {Kirkpatrick},
  {Klaassen}, {Klingemann}, {Klintworth}, {Knapp}, {Knight}, {Knollenberg},
  {Knutsen}, {Koehler}, {Koekemoer}, {Kofler}, {Kontson}, {Kovacs},
  {Kozhurina-Platais}, {Krause}, {Kriss}, {Krist}, {Kristoffersen}, {Krogel},
  {Krueger}, {Kulp}, {Kumari}, {Kwan}, {Kyprianou}, {Gadiano Labador},
  {Labiano}, {Lafreni{\`e}re}, {Lagage}, {Laidler}, {Laine}, {Laird}, {Lajoie},
  {Lallo}, {Yen Lam}, {LaMassa}, {Lambros}, {Lampenfield}, {Lander}, {Hutton
  Langston}, {Larson}, {Larson}, {LaVerghetta}, {Law}, {Lawrence}, {Lee},
  {Lee}, {Lee}, {Leisenring}, {Dunlap Leveille}, {Levenson}, {Levi}, {Levine},
  {Lewis}, {Lewis}, {Lewis}, {Libralato}, {Lidon}, {Liebrecht}, {Lightsey},
  {Lilly}, {Lim}, {Lian Lim}, {Ling}, {Link}, {Link}, {Lipinski}, {Liu}, {Lo},
  {Lobmeyer}, {Logue}, {Long}, {Long}, {Long}, {Long}, {L{\'o}pez-Caniego},
  {Lotz}, {Love-Pruitt}, {Lubskiy}, {Luers}, {Luetgens}, {Luevano}, {Flores
  Lui}, {Lund}, {Lundquist}, {Lunine}, {L{\"u}tzgendorf}, {Lynch}, {MacDonald},
  {MacDonald}, {Macias}, {Macklis}, {Maghami}, {Maharaja}, {Maiolino},
  {Makrygiannis}, {Giri Malla}, {Malumuth}, {Manjavacas}, {Marini}, {Marrione},
  {Marston}, {Martel}, {Martin}, {Martin}, {Martinez}, {Maschmann}, {Masci},
  {Masetti}, {Maszkiewicz}, {Matthews}, {Matuskey}, {McBrayer}, {McCarthy},
  {McCaughrean}, {McClare}, {McClare}, {McCloskey}, {McClurg}, {McCoy},
  {McElwain}, {McGregor}, {McGuffey}, {McKay}, {McKenzie}, {McLean},
  {McMaster}, {McNeil}, {De Meester}, {Mehalick}, {Meixner}, {Mel{\'e}ndez},
  {Menzel}, {Menzel}, {Merz}, {Mesterharm}, {Meyer}, {Meyett}, {Meza},
  {Midwinter}, {Milam}, {Miller}, {Miller}, {Miskey}, {Misselt}, {Mitchell},
  {Mohan}, {Montoya}, {Moran}, {Morishita}, {Moro-Mart{\'\i}n}, {Morrison},
  {Morrison}, {Morse}, {Moschos}, {Moseley}, {Mosier}, {Mosner}, {Mountain},
  {Muckenthaler}, {Mueller}, {Mueller}, {Muhiem}, {M{\"u}hlmann}, {Mullally},
  {Mullen}, {Munger}, {Murphy}, {Murray}, {Muzerolle}, {Mycroft}, {Myers},
  {Myers}, {Myers}, {Myers}, {Myrick}, {Nagle}, {Nayak}, {Naylor}, {Neff},
  {Nelan}, {Nella}, {Tuong Nguyen}, {Nguyen}, {Nickson}, {Nidhiry}, {Niedner},
  {Nieto-Santisteban}, {Nikolov}, {Nishisaka}, {Noriega-Crespo}, {Nota},
  {O'Mara}, {Oboryshko}, {O'Brien}, {Ochs}, {Offenberg}, {Ogle}, {Ohl}, {Hamden
  Olmsted}, {Osborne}, {O'Shaughnessy}, {{\"O}stlin}, {O'Sullivan}, {Otor},
  {Ottens}, {Ouellette}, {Outlaw}, {Owens}, {Pacifici}, {Page}, {Paranilam},
  {Park}, {Parrish}, {Paschal}, {Patapis}, {Patel}, {Patrick}, {Pattishall},
  {Paul}, {Paul}, {Pauly}, {Pavlovsky}, {Pe{\~n}a-Guerrero}, {Pedder}, {Peek},
  {Pelham}, {Penanen}, {Perriello}, {Perrin}, {Perrine}, {Perrygo}, {Peslier},
  {Petach}, {Peterson}, {Pfarr}, {Pierson}, {Pietraszkiewicz}, {Pilchen},
  {Pipher}, {Pirzkal}, {Pitman}, {Player}, {Plesha}, {Plitzke}, {Pohner},
  {Konstantin Poletis}, {Pollizzi}, {Polster}, {Pontius}, {Pontoppidan},
  {Porges}, {Potter}, {Prescott}, {Proffitt}, {Pueyo}, {Aracely Quispe Neira},
  {Radich}, {Rager}, {Rameau}, {Ramey}, {Ramos Alarcon}, {Rampini}, {Rapp},
  {Rashford}, {Rauscher}, {Ravindranath}, {Rawle}, {Rawlings}, {Ray}, {Regan},
  {Rehm}, {Rehm}, {Reid}, {Reis}, {Renk}, {Reoch}, {Ressler}, {Rest},
  {Reynolds}, {Richon}, {Richon}, {Ridgaway}, {Riedel}, {Rieke}, {Rieke},
  {Rifelli}, {Rigby}, {Riggs}, {Ringel}, {Ritchie}, {Rix}, {Robberto},
  {Robinson}, {Robinson}, {Rock}, {Rodriguez}, {Rodr{\'\i}guez del Pino},
  {Roellig}, {Rohrbach}, {Roman}, {Romelfanger}, {Romo}, {Rosales}, {Rose},
  {Roteliuk}, {Roth}, {Quinn Rothwell}, {Rouzaud}, {Rowe}, {Rowlands}, {Roy},
  {Royer}, {Rui}, {Rumler}, {Rumpl}, {Russ}, {Ryan}, {Ryan}, {Saad}, {Sabata},
  {Sabatino}, {Sabbi}, {Sabelhaus}, {Sabia}, {Sahu}, {Saif}, {Salvignol},
  {Samara-Ratna}, {Samuelson}, {Sanders}, {Sappington}, {Sargent}, {Sauer},
  {Savadkin}, {Sawicki}, {Schappell}, {Scheffer}, {Scheithauer}, {Scherer},
  {Schiff}, {Schlawin}, {Schmeitzky}, {Schmitz}, {Schmude}, {Schneider},
  {Schreiber}, {Schroeven-Deceuninck}, {Schultz}, {Schwab}, {Schwartz},
  {Scoccimarro}, {Scott}, {Scott}, {Seaton}, {Seely}, {Seery}, {Seidleck},
  {Sembach}, {Shanahan}, {Shaughnessy}, {Shaw}, {Shay}, {Sheehan}, {Sheth},
  {Shih}, {Shivaei}, {Siegel}, {Sienkiewicz}, {Simmons}, {Simon}, {Sirianni},
  {Sivaramakrishnan}, {Slade}, {Sloan}, {Slocum}, {Slowinski}, {Smith},
  {Smith}, {Smith}, {Smith}, {Smith}, {Smith}, {Smolik}, {Soderblom}, {Sohn},
  {Sokol}, {Sonneborn}, {Sontag}, {Sooy}, {Soummer}, {Southwood}, {Spain},
  {Sparmo}, {Speer}, {Spencer}, {Sprofera}, {Stallcup}, {Stanley},
  {Stansberry}, {Stark}, {Starr}, {Stassi}, {Steck}, {Steeley}, {Stephens},
  {Stephenson}, {Stewart}, {Stiavelli}, {Stockman}, {Strada}, {Straughn},
  {Streetman}, {Strickland}, {Strobele}, {Stuhlinger}, {Stys}, {Such},
  {Sukhatme}, {Sullivan}, {Sullivan}, {Sumner}, {Sun}, {Sunnquist}, {Swade},
  {Swam}, {Swenton}, {Swoish}, {In Tam Litten}, {Tamas}, {Tao}, {Taylor},
  {Taylor}, {te Plate}, {Van Tea}, {Teague}, {Telfer}, {Temim}, {Texter},
  {Thatte}, {Thompson}, {Thompson}, {Thomson}, {Thronson}, {Tierney},
  {Tikkanen}, {Tinnin}, {Tippet}, {Todd}, {Tran}, {Trauger}, {Trejo}, {Truong},
  {Tsukamoto}, {Tufail}, {Tumlinson}, {Tustain}, {Tyra}, {Ubeda}, {Underwood},
  {Uzzo}, {Vaclavik}, {Valenduc}, {Valenti}, {Van Campen}, {van de Wetering},
  {Van Der Marel}, {van Haarlem}, {Vandenbussche}, {Vanterpool}, {Vernoy},
  {Bego{\~n}a Vila Costas}, {Volk}, {Voorzaat}, {Voyton}, {Vydra}, {Waddy},
  {Waelkens}, {Wahlgren}, {Walker}, {Wander}, {Warfield}, {Warner}, {Wasiak},
  {Wasiak}, {Wehner}, {Weiler}, {Weilert}, {Weiss}, {Wells}, {Welty}, {Wheate},
  {Wheeler}, {White}, {Whitehouse}, {Whiteleather}, {Whitman}, {Williams},
  {Willmer}, {Willott}, {Willoughby}, {Wilson}, {Wilson}, {Wilson},
  {Windhorst}, {Wislowski}, {Wolfe}, {Wolfe}, {Wolff}, {Wondel}, {Woo},
  {Woods}, {Worden}, {Workman}, {Wright}, {Wu}, {Wu}, {Wun}, {Wymer},
  {Yadetie}, {Yan}, {Yang}, {Yates}, {Yeager}, {Yerger}, {Young}, {Young},
  {Yu}, {Yu}, {Zak}, {Zeidler}, {Zepp}, {Zhou}, {Zincke}, {Zonak}, \&
  {Zondag}}]{Gardner+2023}
{Gardner}, J.~P., {Mather}, J.~C., {Abbott}, R., {et~al.} 2023, arXiv e-prints,
  arXiv:2304.04869

\bibitem[{{Gardner} {et~al.}(2006){Gardner}, {Mather}, {Clampin}, {Doyon},
  {Greenhouse}, {Hammel}, {Hutchings}, {Jakobsen}, {Lilly}, {Long}, {Lunine},
  {McCaughrean}, {Mountain}, {Nella}, {Rieke}, {Rieke}, {Rix}, {Smith},
  {Sonneborn}, {Stiavelli}, {Stockman}, {Windhorst}, \& {Wright}}]{Gardner2006}
{Gardner}, J.~P., {Mather}, J.~C., {Clampin}, M., {et~al.} 2006, \ssr, 123, 485

\bibitem[{{Gelli} {et~al.}(2023){Gelli}, {Salvadori}, {Ferrara}, {Pallottini},
  \& {Carniani}}]{Gelli+2023}
{Gelli}, V., {Salvadori}, S., {Ferrara}, A., {Pallottini}, A., \& {Carniani},
  S. 2023, arXiv e-prints, arXiv:2303.13574

\bibitem[{{Giavalisco} {et~al.}(2004){Giavalisco}, {Ferguson}, {Koekemoer},
  {Dickinson}, {Alexander}, {Bauer}, {Bergeron}, {Biagetti}, {Brandt},
  {Casertano}, {Cesarsky}, {Chatzichristou}, {Conselice}, {Cristiani}, {Da
  Costa}, {Dahlen}, {de Mello}, {Eisenhardt}, {Erben}, {Fall}, {Fassnacht},
  {Fosbury}, {Fruchter}, {Gardner}, {Grogin}, {Hook}, {Hornschemeier}, {Idzi},
  {Jogee}, {Kretchmer}, {Laidler}, {Lee}, {Livio}, {Lucas}, {Madau},
  {Mobasher}, {Moustakas}, {Nonino}, {Padovani}, {Papovich}, {Park},
  {Ravindranath}, {Renzini}, {Richardson}, {Riess}, {Rosati}, {Schirmer},
  {Schreier}, {Somerville}, {Spinrad}, {Stern}, {Stiavelli}, {Strolger},
  {Urry}, {Vandame}, {Williams}, \& {Wolf}}]{giavalisco2004}
{Giavalisco}, M., {Ferguson}, H.~C., {Koekemoer}, A.~M., {et~al.} 2004, \apjl,
  600, L93

\bibitem[{{Gomes} \& {Papaderos}(2017)}]{Gomes2017}
{Gomes}, J.~M. \& {Papaderos}, P. 2017, \aap, 603, A63

\bibitem[{{Gordon} {et~al.}(2003){Gordon}, {Clayton}, {Misselt}, {Landolt}, \&
  {Wolff}}]{Gordon+2003}
{Gordon}, K.~D., {Clayton}, G.~C., {Misselt}, K.~A., {Landolt}, A.~U., \&
  {Wolff}, M.~J. 2003, \apj, 594, 279

\bibitem[{{Jakobsen} {et~al.}(2022){Jakobsen}, {Ferruit}, {Alves de Oliveira},
  {Arribas}, {Bagnasco}, {Barho}, {Beck}, {Birkmann}, {B{\"o}ker}, {Bunker},
  {Charlot}, {de Jong}, {de Marchi}, {Ehrenwinkler}, {Falcolini}, {Fels},
  {Franx}, {Franz}, {Funke}, {Giardino}, {Gnata}, {Holota}, {Honnen}, {Jensen},
  {Jentsch}, {Johnson}, {Jollet}, {Karl}, {Kling}, {K{\"o}hler}, {Kolm},
  {Kumari}, {Lander}, {Lemke}, {L{\'o}pez-Caniego}, {L{\"u}tzgendorf},
  {Maiolino}, {Manjavacas}, {Marston}, {Maschmann}, {Maurer}, {Messerschmidt},
  {Moseley}, {Mosner}, {Mott}, {Muzerolle}, {Pirzkal}, {Pittet}, {Plitzke},
  {Posselt}, {Rapp}, {Rauscher}, {Rawle}, {Rix}, {R{\"o}del}, {Rumler},
  {Sabbi}, {Salvignol}, {Schmid}, {Sirianni}, {Smith}, {Strada}, {te Plate},
  {Valenti}, {Wettemann}, {Wiehe}, {Wiesmayer}, {Willott}, {Wright}, {Zeidler},
  \& {Zincke}}]{jakobsen+2022}
{Jakobsen}, P., {Ferruit}, P., {Alves de Oliveira}, C., {et~al.} 2022, \aap,
  661, A80

\bibitem[{{Ji} \& {Giavalisco}(2022)}]{Ji+2022}
{Ji}, Z. \& {Giavalisco}, M. 2022, \apj, 935, 120

\bibitem[{{Ji} \& {Giavalisco}(2023)}]{Ji+2023}
{Ji}, Z. \& {Giavalisco}, M. 2023, \apj, 943, 54

\bibitem[{{Johnson} {et~al.}(2021){Johnson}, {Leja}, {Conroy}, \&
  {Speagle}}]{johnson+2021}
{Johnson}, B.~D., {Leja}, J., {Conroy}, C., \& {Speagle}, J.~S. 2021, \apjs,
  254, 22

\bibitem[{{Kawata} \& {Gibson}(2003)}]{Kawata&Gibson2003}
{Kawata}, D. \& {Gibson}, B.~K. 2003, \mnras, 346, 135

\bibitem[{{Kennicutt}(1998)}]{kennicutt1998}
{Kennicutt}, Robert~C., J. 1998, \apj, 498, 541

\bibitem[{Kennicutt \& Evans(2012)}]{kennicutt_star_2012}
Kennicutt, R.~C. \& Evans, N.~J. 2012, \araa, 50, 531

\bibitem[{{Koudmani} {et~al.}(2019){Koudmani}, {Sijacki}, {Bourne}, \&
  {Smith}}]{Koudmani2019MNRAS.484.2047K}
{Koudmani}, S., {Sijacki}, D., {Bourne}, M.~A., \& {Smith}, M.~C. 2019, \mnras,
  484, 2047

\bibitem[{{Leja} {et~al.}(2019){Leja}, {Carnall}, {Johnson}, {Conroy}, \&
  {Speagle}}]{leja2019}
{Leja}, J., {Carnall}, A.~C., {Johnson}, B.~D., {Conroy}, C., \& {Speagle},
  J.~S. 2019, \apj, 876, 3

\bibitem[{{Looser} {et~al.}(2023){Looser}, {D'Eugenio}, {Maiolino}, {Witstok},
  {Sandles}, {Curtis-Lake}, {Chevallard}, {Tacchella}, {Johnson}, {Baker},
  {Suess}, {Carniani}, {Ferruit}, {Arribas}, {Bonaventura}, {Bunker},
  {Cameron}, {Charlot}, {Curti}, {de Graaff}, {Maseda}, {Rawle}, {Rix},
  {Rodriguez Del Pino}, {Smit}, {{\"U}bler}, {Willott}, {Alberts}, {Egami},
  {Eisenstein}, {Endsley}, {Hausen}, {Rieke}, {Robertson}, {Shivaei},
  {Williams}, {Boyett}, {Chen}, {Ji}, {Jones}, {Kumari}, {Nelson}, {Perna},
  {Saxena}, \& {Scholtz}}]{Looser+2023}
{Looser}, T.~J., {D'Eugenio}, F., {Maiolino}, R., {et~al.} 2023, arXiv
  e-prints, arXiv:2302.14155

\bibitem[{{Lovell} {et~al.}(2022){Lovell}, {Roper}, {Vijayan}, {Seeyave},
  {Irodotou}, {Wilkins}, {Conselice}, {Fortuni}, {Kuusisto}, {Merlin},
  {Santini}, \& {Thomas}}]{Lovell2022arXiv221107540L}
{Lovell}, C.~C., {Roper}, W., {Vijayan}, A.~P., {et~al.} 2022, arXiv e-prints,
  arXiv:2211.07540

\bibitem[{{Lu} {et~al.}(2023){Lu}, {Zhu}, {Cappellari}, {Li}, {Mao}, \&
  {Xu}}]{Lu2023}
{Lu}, S., {Zhu}, K., {Cappellari}, M., {et~al.} 2023, arXiv e-prints,
  arXiv:2304.11712

\bibitem[{{Ma} {et~al.}(2018){Ma}, {Hopkins}, {Garrison-Kimmel},
  {Faucher-Gigu{\`e}re}, {Quataert}, {Boylan-Kolchin}, {Hayward}, {Feldmann},
  \& {Kere{\v{s}}}}]{Ma2018MNRAS.478.1694M}
{Ma}, X., {Hopkins}, P.~F., {Garrison-Kimmel}, S., {et~al.} 2018, \mnras, 478,
  1694

\bibitem[{{Maheson} {et~al.}(2023){Maheson}, {Maiolino}, {Curti}, {Sanders},
  {Tacchella}, \& {Sandles}}]{Maheson23}
{Maheson}, G., {Maiolino}, R., {Curti}, M., {et~al.} 2023, arXiv e-prints,
  arXiv:2306.00069

\bibitem[{{McLure} {et~al.}(2018){McLure}, {Dunlop}, {Cullen}, {Bourne},
  {Best}, {Khochfar}, {Bowler}, {Biggs}, {Geach}, {Scott}, {Micha{\l}owski},
  {Rujopakarn}, {van Kampen}, {Kirkpatrick}, \& {Pope}}]{McLure2018}
{McLure}, R.~J., {Dunlop}, J.~S., {Cullen}, F., {et~al.} 2018, \mnras, 476,
  3991

\bibitem[{{Noll} {et~al.}(2009){Noll}, {Burgarella}, {Giovannoli}, {Buat},
  {Marcillac}, \& {Mu{\~n}oz-Mateos}}]{Noll2009}
{Noll}, S., {Burgarella}, D., {Giovannoli}, E., {et~al.} 2009, \aap, 507, 1793

\bibitem[{{Pacifici} {et~al.}(2016){Pacifici}, {Kassin}, {Weiner}, {Holden},
  {Gardner}, {Faber}, {Ferguson}, {Koo}, {Primack}, {Bell}, {Dekel}, {Gawiser},
  {Giavalisco}, {Rafelski}, {Simons}, {Barro}, {Croton}, {Dav{\'e}}, {Fontana},
  {Grogin}, {Koekemoer}, {Lee}, {Salmon}, {Somerville}, \&
  {Behroozi}}]{pacifici+2016}
{Pacifici}, C., {Kassin}, S.~A., {Weiner}, B.~J., {et~al.} 2016, \apj, 832, 79

\bibitem[{{Pannella} {et~al.}(2009){Pannella}, {Carilli}, {Daddi}, {McCracken},
  {Owen}, {Renzini}, {Strazzullo}, {Civano}, {Koekemoer}, {Schinnerer},
  {Scoville}, {Smol{\v{c}}i{\'c}}, {Taniguchi}, {Aussel}, {Kneib}, {Ilbert},
  {Mellier}, {Salvato}, {Thompson}, \& {Willott}}]{Pannella2009}
{Pannella}, M., {Carilli}, C.~L., {Daddi}, E., {et~al.} 2009, \apjl, 698, L116

\bibitem[{{Pappalardo} {et~al.}(2021){Pappalardo}, {Cardoso}, {Michel Gomes},
  {Papaderos}, {Afonso}, {Breda}, {Humphrey}, {Scott}, {Amarantidis}, {Matute},
  {Carvajal}, {Lorenzoni}, {Lagos}, {Paulino-Afonso}, \&
  {Miranda}}]{Pappalardo2021}
{Pappalardo}, C., {Cardoso}, L. S.~M., {Michel Gomes}, J., {et~al.} 2021, \aap,
  651, A99

\bibitem[{{Peng} {et~al.}(2015){Peng}, {Maiolino}, \&
  {Cochrane}}]{Peng2015Natur.521..192P}
{Peng}, Y., {Maiolino}, R., \& {Cochrane}, R. 2015, \nat, 521, 192

\bibitem[{{Peng} {et~al.}(2010){Peng}, {Lilly}, {Kova{\v c}}, {Bolzonella},
  {Pozzetti}, {Renzini}, {Zamorani}, {Ilbert}, {Knobel}, {Iovino}, {Maier},
  {Cucciati}, {Tasca}, {Carollo}, {Silverman}, {Kampczyk}, {de Ravel},
  {Sanders}, {Scoville}, {Contini}, {Mainieri}, {Scodeggio}, {Kneib}, {Le
  F{\`e}vre}, {Bardelli}, {Bongiorno}, {Caputi}, {Coppa}, {de la Torre},
  {Franzetti}, {Garilli}, {Lamareille}, {Le Borgne}, {Le Brun}, {Mignoli},
  {Perez Montero}, {Pello}, {Ricciardelli}, {Tanaka}, {Tresse}, {Vergani},
  {Welikala}, {Zucca}, {Oesch}, {Abbas}, {Barnes}, {Bordoloi}, {Bottini},
  {Cappi}, {Cassata}, {Cimatti}, {Fumana}, {Hasinger}, {Koekemoer},
  {Leauthaud}, {Maccagni}, {Marinoni}, {McCracken}, {Memeo}, {Meneux}, {Nair},
  {Porciani}, {Presotto}, \& {Scaramella}}]{peng+2010}
{Peng}, Y.-j., {Lilly}, S.~J., {Kova{\v c}}, K., {et~al.} 2010, \apj, 721, 193

\bibitem[{{Rieke} {et~al.}(2023){Rieke}, {Kelly}, {Misselt}, {Stansberry},
  {Boyer}, {Beatty}, {Egami}, {Florian}, {Greene}, {Hainline}, {Leisenring},
  {Roellig}, {Schlawin}, {Sun}, {Tinnin}, {Williams}, {Willmer}, {Wilson},
  {Clark}, {Rohrbach}, {Brooks}, {Canipe}, {Correnti}, {DiFelice}, {Gennaro},
  {Girard}, {Hartig}, {Hilbert}, {Koekemoer}, {Nikolov}, {Pirzkal}, {Rest},
  {Robberto}, {Sunnquist}, {Telfer}, {Wu}, {Ferry}, {Lewis}, {Baum},
  {Beichman}, {Doyon}, {Dressler}, {Eisenstein}, {Ferrarese}, {Hodapp},
  {Horner}, {Jaffe}, {Johnstone}, {Krist}, {Martin}, {McCarthy}, {Meyer},
  {Rieke}, {Trauger}, \& {Young}}]{Rieke2023}
{Rieke}, M.~J., {Kelly}, D.~M., {Misselt}, K., {et~al.} 2023, \pasp, 135,
  028001

\bibitem[{{Robotham} {et~al.}(2020){Robotham}, {Bellstedt}, {Lagos}, {Thorne},
  {Davies}, {Driver}, \& {Bravo}}]{Robotham2020}
{Robotham}, A.~S.~G., {Bellstedt}, S., {Lagos}, C. d.~P., {et~al.} 2020,
  \mnras, 495, 905

\bibitem[{{Sandles} {et~al.}(2022){Sandles}, {Curtis-Lake}, {Charlot},
  {Chevallard}, \& {Maiolino}}]{Sandles2022MNRAS.515.2951S}
{Sandles}, L., {Curtis-Lake}, E., {Charlot}, S., {Chevallard}, J., \&
  {Maiolino}, R. 2022, \mnras, 515, 2951

\bibitem[{{Shapley} {et~al.}(2023){Shapley}, {Sanders}, {Reddy}, {Topping}, \&
  {Brammer}}]{shapley2023}
{Shapley}, A.~E., {Sanders}, R.~L., {Reddy}, N.~A., {Topping}, M.~W., \&
  {Brammer}, G.~B. 2023, arXiv e-prints, arXiv:2301.03241

\bibitem[{{Shapley} {et~al.}(2022){Shapley}, {Sanders}, {Salim}, {Reddy},
  {Kriek}, {Mobasher}, {Coil}, {Siana}, {Price}, {Shivaei}, {Dunlop}, {McLure},
  \& {Cullen}}]{Shapley2022}
{Shapley}, A.~E., {Sanders}, R.~L., {Salim}, S., {et~al.} 2022, \apj, 926, 145

\bibitem[{{Shivaei} {et~al.}(2015){Shivaei}, {Reddy}, {Steidel}, \&
  {Shapley}}]{Shivaei2015}
{Shivaei}, I., {Reddy}, N.~A., {Steidel}, C.~C., \& {Shapley}, A.~E. 2015,
  \apj, 804, 149

\bibitem[{{Smit} {et~al.}(2015){Smit}, {Bouwens}, {Franx}, {Oesch}, {Ashby},
  {Willner}, {Labb{\'e}}, {Holwerda}, {Fazio}, \& {Huang}}]{Smit+2015}
{Smit}, R., {Bouwens}, R.~J., {Franx}, M., {et~al.} 2015, \apj, 801, 122

\bibitem[{{Smit} {et~al.}(2014){Smit}, {Bouwens}, {Labb{\'e}}, {Zheng},
  {Bradley}, {Donahue}, {Lemze}, {Moustakas}, {Umetsu}, {Zitrin}, {Coe},
  {Postman}, {Gonzalez}, {Bartelmann}, {Ben{\'\i}tez}, {Broadhurst}, {Ford},
  {Grillo}, {Infante}, {Jimenez-Teja}, {Jouvel}, {Kelson}, {Lahav}, {Maoz},
  {Medezinski}, {Melchior}, {Meneghetti}, {Merten}, {Molino}, {Moustakas},
  {Nonino}, {Rosati}, \& {Seitz}}]{Smit+2014}
{Smit}, R., {Bouwens}, R.~J., {Labb{\'e}}, I., {et~al.} 2014, \apj, 784, 58

\bibitem[{{Speagle} {et~al.}(2014){Speagle}, {Steinhardt}, {Capak}, \&
  {Silverman}}]{speagle+2014}
{Speagle}, J.~S., {Steinhardt}, C.~L., {Capak}, P.~L., \& {Silverman}, J.~D.
  2014, \apjs, 214, 15

\bibitem[{{Strait} {et~al.}(2023){Strait}, {Brammer}, {Muzzin}, {Dezprez},
  {Asada}, {Abraham}, {Brada{\v{c}}}, {Iyer}, {Martis}, {Mowla}, {Noirot},
  {Sarrouh}, {Sawicki}, {Willott}, {Gould}, {Grindlay}, {Matharu}, \&
  {Rihtar{\v{s}}i{\v{c}}}}]{Strait2023}
{Strait}, V., {Brammer}, G., {Muzzin}, A., {et~al.} 2023, arXiv e-prints,
  arXiv:2303.11349

\bibitem[{{Sun} {et~al.}(2023){Sun}, {Faucher-Gigu{\`e}re}, {Hayward}, \&
  {Shen}}]{Sun+2023}
{Sun}, G., {Faucher-Gigu{\`e}re}, C.-A., {Hayward}, C.~C., \& {Shen}, X. 2023,
  arXiv e-prints, arXiv:2305.02713

\bibitem[{{Tacchella} {et~al.}(2022){Tacchella}, {Conroy}, {Faber}, {Johnson},
  {Leja}, {Barro}, {Cunningham}, {Deason}, {Guhathakurta}, {Guo}, {Hernquist},
  {Koo}, {McKinnon}, {Rockosi}, {Speagle}, {van Dokkum}, \&
  {Yesuf}}]{Tacchella2022ApJ...926..134T}
{Tacchella}, S., {Conroy}, C., {Faber}, S.~M., {et~al.} 2022, \apj, 926, 134

\bibitem[{{Tacchella} {et~al.}(2016){Tacchella}, {Dekel}, {Carollo},
  {Ceverino}, {DeGraf}, {Lapiner}, {Mandelker}, \& {Primack
  Joel}}]{Tacchella2016}
{Tacchella}, S., {Dekel}, A., {Carollo}, C.~M., {et~al.} 2016, \mnras, 457,
  2790

\bibitem[{{Tacchella} {et~al.}(2020{\natexlab{a}}){Tacchella}, {Forbes}, \&
  {Caplar}}]{Tacchella2020}
{Tacchella}, S., {Forbes}, J.~C., \& {Caplar}, N. 2020{\natexlab{a}}, \mnras,
  497, 698

\bibitem[{{Tacchella} {et~al.}(2020{\natexlab{b}}){Tacchella}, {Forbes}, \&
  {Caplar}}]{Tacchella2020MNRAS.497..698T}
{Tacchella}, S., {Forbes}, J.~C., \& {Caplar}, N. 2020{\natexlab{b}}, \mnras,
  497, 698

\bibitem[{{Trussler} {et~al.}(2020){Trussler}, {Maiolino}, {Maraston}, {Peng},
  {Thomas}, {Goddard}, \& {Lian}}]{Trussler2020}
{Trussler}, J., {Maiolino}, R., {Maraston}, C., {et~al.} 2020, \mnras, 491,
  5406

\bibitem[{{Trussler} {et~al.}(2021){Trussler}, {Maiolino}, {Maraston}, {Peng},
  {Thomas}, {Goddard}, \& {Lian}}]{Trussler2021MNRAS.500.4469T}
{Trussler}, J., {Maiolino}, R., {Maraston}, C., {et~al.} 2021, \mnras, 500,
  4469

\bibitem[{{Vidal-Garc{\'\i}a} {et~al.}(2017){Vidal-Garc{\'\i}a}, {Charlot},
  {Bruzual}, \& {Hubeny}}]{vidal_garcia_2017}
{Vidal-Garc{\'\i}a}, A., {Charlot}, S., {Bruzual}, G., \& {Hubeny}, I. 2017,
  \mnras, 470, 3532

\bibitem[{{Wang} {et~al.}(2019){Wang}, {Lilly}, {Pezzulli}, \&
  {Matthee}}]{wang+2019}
{Wang}, E., {Lilly}, S.~J., {Pezzulli}, G., \& {Matthee}, J. 2019, \apj, 877,
  132

\bibitem[{{Wang} {et~al.}(2016){Wang}, {Elbaz}, {Daddi}, {Finoguenov}, {Liu},
  {Schreiber}, {Mart{\'\i}n}, {Strazzullo}, {Valentino}, {van der Burg},
  {Zanella}, {Ciesla}, {Gobat}, {Le Brun}, {Pannella}, {Sargent}, {Shu}, {Tan},
  {Cappelluti}, \& {Li}}]{Wang2016}
{Wang}, T., {Elbaz}, D., {Daddi}, E., {et~al.} 2016, \apj, 828, 56

\bibitem[{{Weisz} {et~al.}(2012{\natexlab{a}}){Weisz}, {Johnson}, {Johnson},
  {Skillman}, {Lee}, {Kennicutt}, {Calzetti}, {van Zee}, {Bothwell},
  {Dalcanton}, {Dale}, \& {Williams}}]{Weisz2012}
{Weisz}, D.~R., {Johnson}, B.~D., {Johnson}, L.~C., {et~al.}
  2012{\natexlab{a}}, \apj, 744, 44

\bibitem[{{Weisz} {et~al.}(2012{\natexlab{b}}){Weisz}, {Johnson}, {Johnson},
  {Skillman}, {Lee}, {Kennicutt}, {Calzetti}, {van Zee}, {Bothwell},
  {Dalcanton}, {Dale}, \& {Williams}}]{weisz+2012}
{Weisz}, D.~R., {Johnson}, B.~D., {Johnson}, L.~C., {et~al.}
  2012{\natexlab{b}}, \apj, 744, 44

\bibitem[{{Whitaker} {et~al.}(2017){Whitaker}, {Pope}, {Cybulski}, {Casey},
  {Popping}, \& {Yun}}]{Whitaker2017}
{Whitaker}, K.~E., {Pope}, A., {Cybulski}, R., {et~al.} 2017, \apj, 850, 208

\bibitem[{{Zhu} {et~al.}(2023{\natexlab{a}}){Zhu}, {Lu}, {Cappellari}, {Li},
  {Mao}, \& {Gao}}]{Zhu2023b}
{Zhu}, K., {Lu}, S., {Cappellari}, M., {et~al.} 2023{\natexlab{a}}, arXiv
  e-prints, arXiv:2304.11714

\bibitem[{{Zhu} {et~al.}(2023{\natexlab{b}}){Zhu}, {Lu}, {Cappellari}, {Li},
  {Mao}, \& {Gao}}]{Zhu2023}
{Zhu}, K., {Lu}, S., {Cappellari}, M., {et~al.} 2023{\natexlab{b}}, \mnras,
  522, 6326

\end{thebibliography}

\begin{appendix}

\section{\ppxf stellar population grid fitting example}

\begin{figure}[h!]
   \centering
\includegraphics[width=1.\columnwidth]{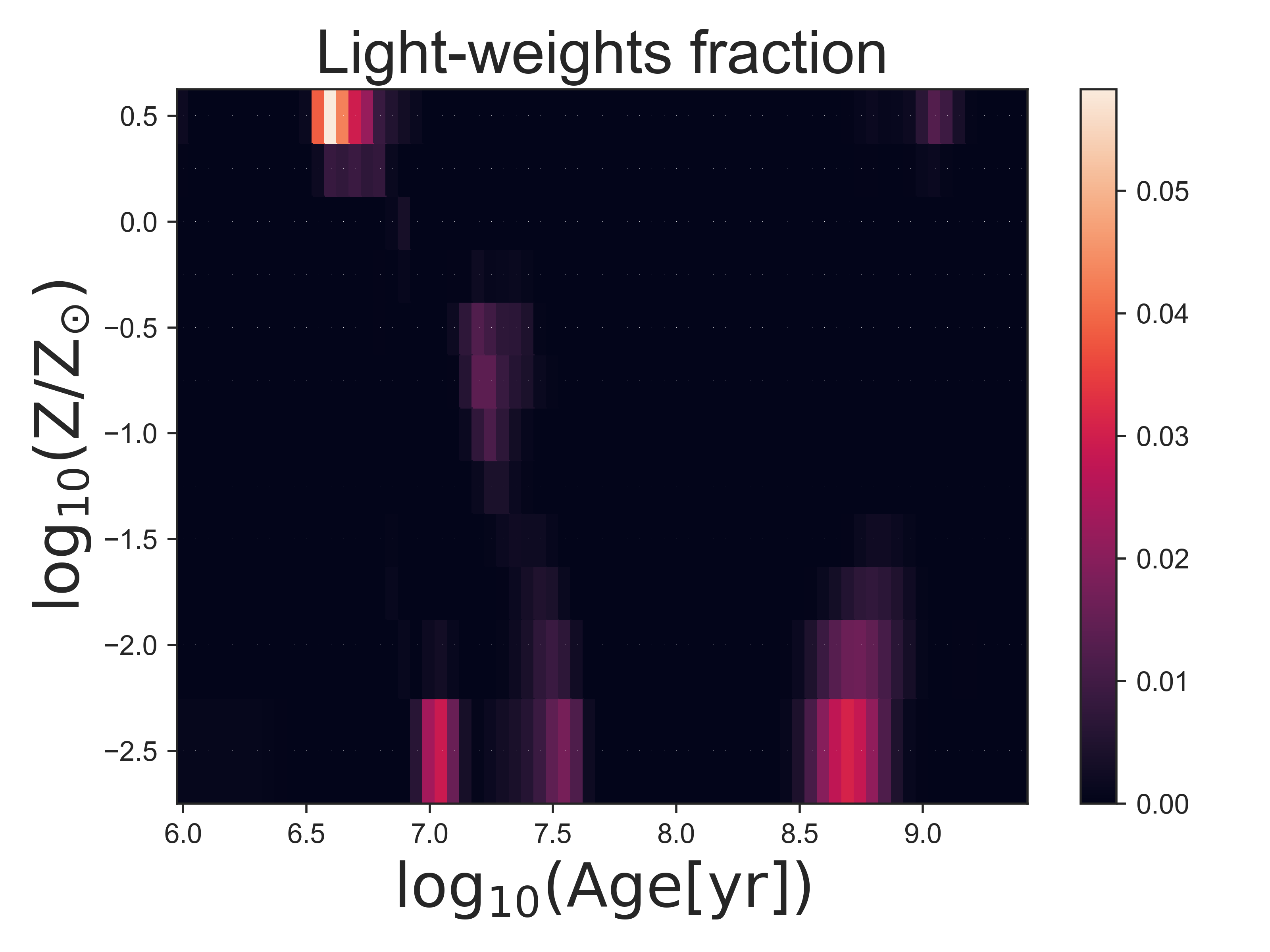}
\caption{Example: \ppxf stellar population inference from the spectrum presented in Fig.~\ref{fig:spectrum_005329}. The fitted light-weighted SSP age-metallicity grid.}\label{ppxf_SFH_example_2D}
\end{figure}

\begin{figure}
\includegraphics[width=1.\columnwidth]{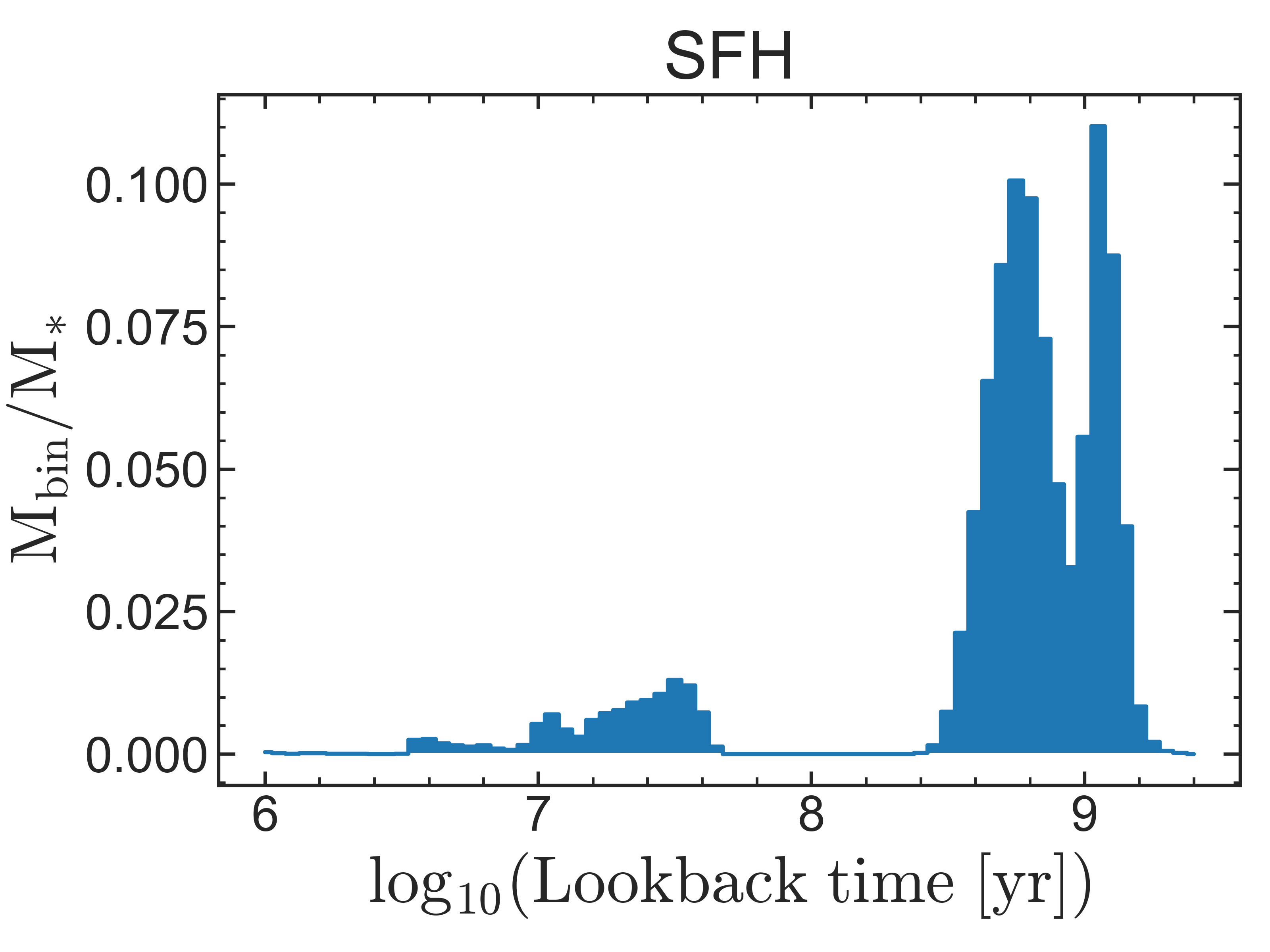}
    \caption{Example: \ppxf SFH inferred from the spectrum presented in Fig.~\ref{fig:spectrum_005329}. The 2D weight-grid (shown in Fig.\ref{ppxf_SFH_example_2D}) converted to a mass-weighted SFH, where the ages of the SSP-templates are equivalent to the star formation activity at that look-back time. The SFH is normalized by the total stellar mass of the galaxy \Mstar.} 
    \label{ppxf_SFH_example_1D}
\end{figure}

To show an illustrative example of our SFH inference methodology based on \ppxf, we show the results of the fitting of the spectrum shown in Fig.~\ref{fig:spectrum_005329}. In Fig.~\ref{ppxf_SFH_example_2D} the light-weighted 2D grid of SSP-weights is presented, where the age of the SSP templates is given on the x-axis, and the stellar metallicity on the y-axis, respectively. Fig.~\ref{ppxf_SFH_example_1D} shows the conversion of these weights into a non-parametric mass-weighted SFH.

\section{Comparison between \ppxf and \beagle stellar mass}\label{comp_masses}

\begin{figure}
    \centering
    \includegraphics[width=\columnwidth]{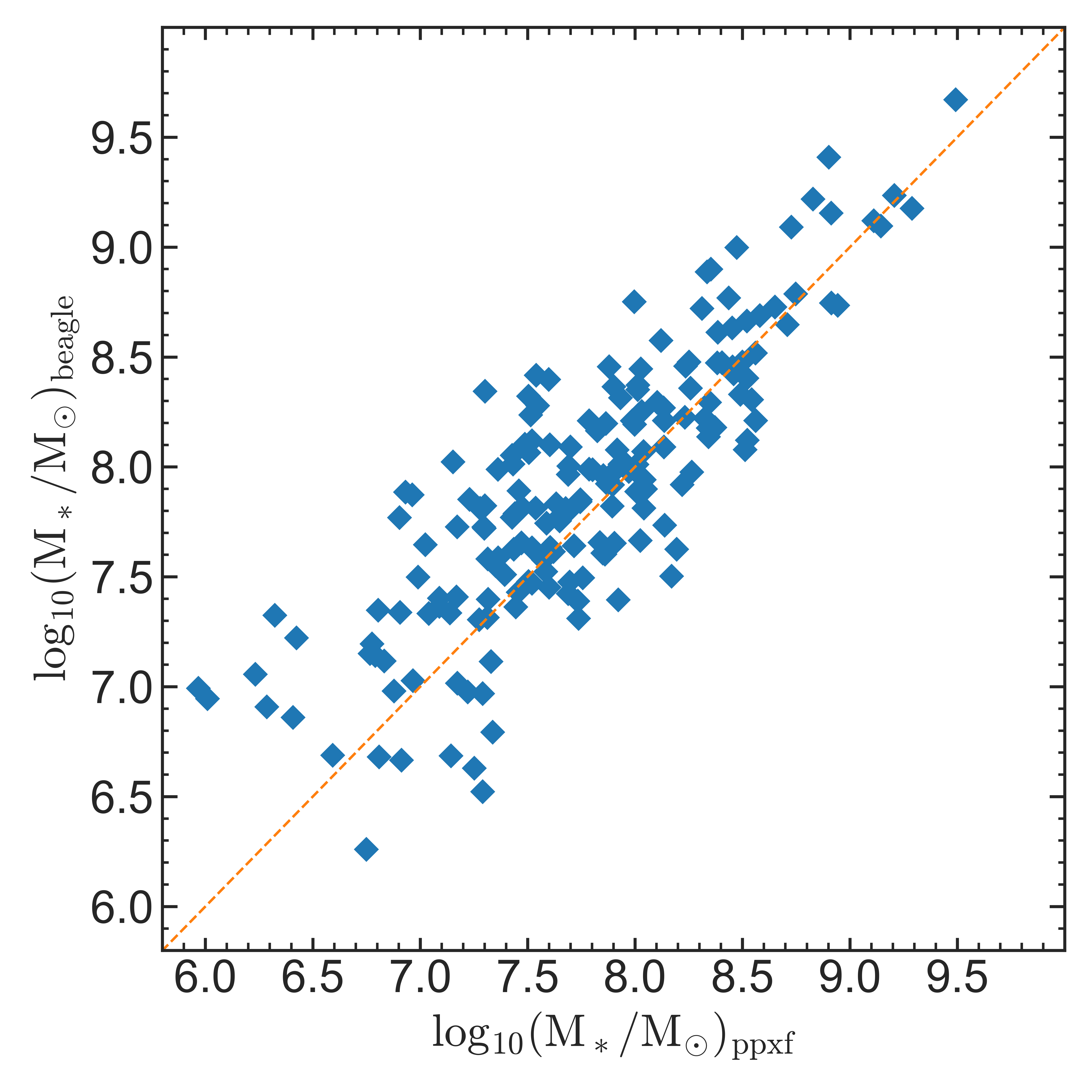}
    \caption{Comparison between the stellar masses inferred by \ppxf and \beagle. The orange line indicates the 1:1 line. The masses show a strong correlation, with an RMS-scatter of ~0.2 dex, but exhibit an offset of ~0.2 dex.}
    \label{fig:comp_mass}
\end{figure}

In Fig.~\ref{fig:comp_mass}, we present a comparison between the stellar masses inferred by \ppxf, see details in section \ref{sec:mass_sfr}, and those inferred by the Bayesian inference code \beagle \citep{chevallard_beagle_2016}. The \beagle-derived masses are computed assuming a parametric delayed exponential SFH combined with a 10 Myr burst; and adopting an updated version of the BC03 stellar population model library \citep{bruzual_stellar_2003}, as described in \cite{vidal_garcia_2017}. More details on the \beagle stellar masses for the JADES/HST-DEEP sample are given in \citet{Curti2023} and Chevallard et al. (in~prep.).
Overall, we find a strong correlation between the stellar masses inferred by the two codes, with a linear fit of $\mathrm{log_{10}(M_{*,\beagle}) = (0.76 \pm 0.04)}\times \mathrm{(log_{10}(M_{*,\ppxf}) - 8.0) }$$ + (8.12 \pm 0.02)$ and a RMS-scatter of ~0.2 dex. However, we note a ~0.2 dex offset in overall normalisation. The larger masses inferred by \beagle compared to \ppxf likely stem from the different modelling approaches.
%, based on non-parametric versus parametric SFH approaches, might play a role.

\end{appendix}

\end{document}